\documentclass[aps,pra,showpacs,twocolumn,amsmath,amssymb,superscriptaddress,nofootinbib]{revtex4}
\usepackage{graphicx}
\usepackage{bm}
\usepackage{amssymb}
\usepackage{amsmath}
\usepackage{latexsym}
\usepackage{color}
\newcommand{\vc}[1]{\mbox{\boldmath $#1$}} 
\newcommand{\ind}[1]{_{#1}}    
\newcommand{\indrm}[1]{_{\mathrm {#1}}}    
\newcommand{\hkl}{_{H}}    

\newcommand{\ehs}{\epsilon_{\hkl}^{({\mathrm s})}}   
\newcommand{\deis}[1]{\Delta E_{\indrm{#1}}^{({\mathrm s})}}   
\newcommand{\dais}[1]{\Delta \theta_{\indrm{#1}}^{({\mathrm s})}}   
\newcommand{\dei}[1]{\Delta E_{\indrm{#1}}}   
\newcommand{\dai}[1]{\Delta \theta_{\indrm{#1}}}   
\newcommand{\CT}{Czerny-Turner}   
\newcommand{\dirate}{{\mathcal D}}   
\newcommand{\sgn}{s}   

\newcommand{\fspace}{P}   
\newcommand{\thinlens}{L}   
\newcommand{\crystal}{C}   
\newcommand{\crystalsp}{K}   
   
\newcommand{\spectro}{S}   
\newcommand{\focusimgmono}{M}   
\newcommand{\focus}{F}   
\newcommand{\dcomm}[1]{b_{\cup_{\ind{#1}}}}   
\newcommand{\acomm}[1]{1/b_{\cup_{\ind{#1}}}}
\newcommand{\bcomm}[1]{B_{\cup_{\ind{#1}}}}      
\newcommand{\gcomm}[1]{G_{\cup_{\ind{#1}}}}      
\newcommand{\fcomm}[1]{\dirate_{\cup_{\ind{#1}}}}
\newcommand{\fmo}{I}   
\newcommand{\fmt}{II}   
\newcommand{\sz}{x_{\indrm{b}}}   
\newcommand{\elmt}{e}   

\begin{document}  
\title{Theory of  angular dispersive  imaging hard x-ray spectrographs} 
\author{Yuri  Shvyd'ko}
\email{shvydko@aps.anl.gov} 
\affiliation{Advanced Photon  Source, Argonne National Laboratory, Argonne, Illinois 60439, USA}
\begin{abstract} 
  A spectrograph is an optical instrument that disperses photons of
  different energies into distinct directions and space locations, and
  images photon spectra on a position-sensitive
  detector. Spectrographs consist of collimating, angular dispersive,
  and focusing optical elements.  Bragg reflecting crystals arranged
  in an asymmetric scattering geometry are used as the dispersing
  elements.  A ray-transfer matrix technique is applied to propagate
  x-rays through the optical elements. Several optical designs of hard
  x-ray spectrographs are proposed and their performance is
  analyzed. Spectrographs with an energy resolution of 0.1~meV and a
  spectral window of imaging up to a few tens of meVs are shown to be
  feasible for inelastic x-ray scattering (IXS) spectroscopy
  applications. In another example, a spectrograph with a 1-meV
  spectral resolution and $85$-meV spectral window of imaging is
  considered for Cu K-edge resonant IXS (RIXS).

\end{abstract}

\pacs{41.50.+h, 07.85.Nc, 61.10.-i, 78.70.Ck}
\maketitle

\section{Introduction}
\label{intro}

\begin{figure}[t!]
\setlength{\unitlength}{\textwidth}
\begin{picture}(1,0.455)(0,0)
\put(0.,0.00){\includegraphics[width=0.50\textwidth]{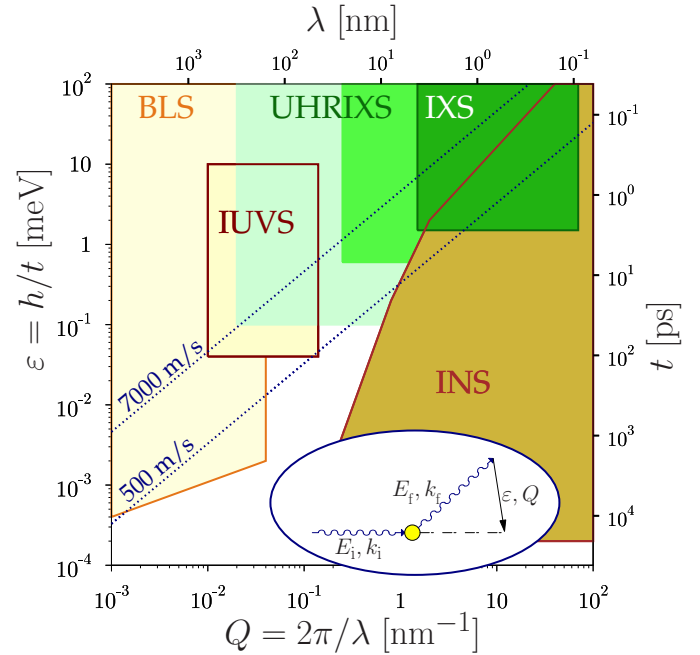}}
\end{picture}
\caption{(Color online) Time-length ($t-\lambda$) and relevant
  energy-momentum ($\varepsilon\!-\!Q$) space of excitations in
  condensed matter and how it is accessed by different inelastic
  scattering probes: neutron (INS), x-ray (IXS), ultraviolet (IUVS),
  and Brillouin (BLS). The ultra-high-resolution IXS spectrometer
  presented in Ref.~\cite{SSS14} entered the previously inaccessible
  region marked in green. The novel capabilities discussed in the
  present paper will enable IXS experiments with even higher
  resolution, 0.1-meV and 0.01-nm$^{-1}$, in the region marked in
  light green, and will close completely the existing gap between the
  high-frequency and low-frequency probes. The energy
  $\varepsilon=E_{\indrm{f}}\!-\!E_{\indrm{i}}$ and the momentum
  $\vc{Q}=\vc{k}_{\indrm{f}}\!-\!\vc{k}_{\indrm{i}}$ transfers from
  initial to final photon/neutron states are measured in inelastic
  scattering experiments, schematically shown in the oval inset.}
\label{fig000}
\end{figure}

Ultra-fast dynamics in condensed matter in a picosecond (ps) to a
100-ps regime on atomic- to meso-scales is still inaccessible for
studies using any known experimental probe. A gap remains in
experimental capabilities between the low-frequency (visible and
ultraviolet light) and high-frequency (x-rays and neutrons) inelastic
scattering techniques.  Figure~\ref{fig000} shows how the time-length
space or the relevant energy-momentum space of excitations in
condensed matter is accessed by different inelastic scattering probes:
neutron (INS), x-ray (IXS), ultraviolet (IUVS), and Brillouin (BLS);
as well as how the remaining gap could be closed by enhancing
inelastic x-ray scattering capabilities.  Ultra-high-resolution IXS
(UHRIXS) has the potential to enter the unexplored dynamic range of
excitations in condensed matter. This would, however, require
achieving a very high spectral resolution on the order of 0.1~meV, and
momentum transfer resolution around 0.01~nm$^{-1}$ (light green area
in Fig.~\ref{fig000}).  In approaching this goal, a novel IXS
spectrometer has been demonstrated recently \cite{SSS14}; the spectral
resolution improved from 1.5~meV to 0.6~meV, the momentum transfer
resolution improved from 1~nm$^{-1}$ to 0.25~nm$^{-1}$ (dark-green and
green areas in Fig.~\ref{fig000}, respectively), and the spectral
contrast improved by an order of magnitude compared to the traditional
IXS spectrometers \cite{BDP87,SRK95,MBKRSV96,Baron1,SAB01,SSD11}.  The
gap became narrower, but did not close.

The outstanding problems in the condensed matter physics, such as the
nature of the liquid to glass transitions, have yet to be fully
addressed.  Here we propose an approach of how this problem could be
solved, and how UHRIXS spectrometers could become efficient imaging
optical devices.  This approach is a further development of the
proposal presented in \cite{Shv11,Shvydko12,SSM13}.

\begin{figure*}[t!]
\setlength{\unitlength}{\textwidth}
\begin{picture}(1,0.30)(0,0)
\put(0.025,0.00){\includegraphics[width=0.95\textwidth]{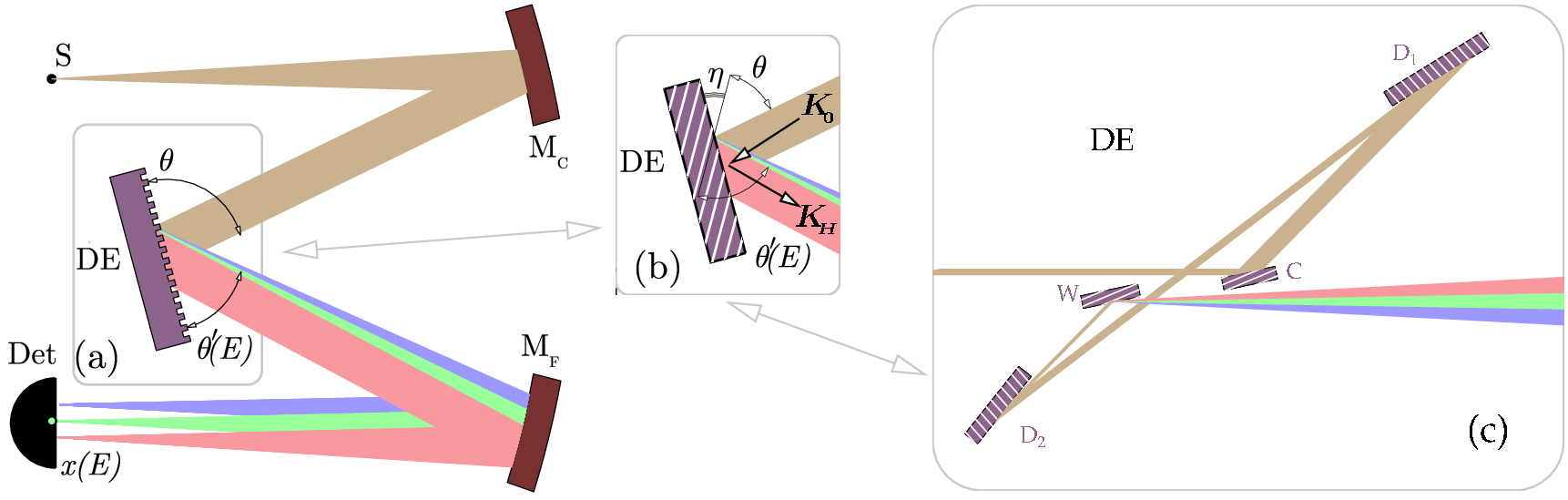}}
\end{picture}
\caption{(Color online) Scheme of the \CT\ type spectrograph
  \cite{CT30} with (a) a diffraction grating, or (b) a crystal in
  asymmetric x-ray Bragg diffraction as dispersing element (DE). Other
  components include radiation source S, collimating and focusing
  mirrors M$_{\indrm{C}}$ and M$_{\indrm{F}}$, and position sensitive
  detector Det. (c) Multi-crystal multi-reflection CDDW optic, an
  example of a hard x-ray ``diffraction grating'' (DE element) with
  enhanced dispersion rate, suitable for hard x-ray spectrographs
  \cite{SSM13}.}
\label{fig001}
\end{figure*}

In a typical IXS experiment \cite{Sinn01}, x-rays incident on a sample
are monochromatized to a very small bandwidth corresponding to a
desired energy resolution. The spectral analysis of photons scattered
from a sample is performed by an x-ray analyzer, featuring the same
spectral bandwidth and acting like a spectral slit.
Monochromatization from approximately a $100$-eV to a $1$-meV
bandwidth results in a dramatic reduction of the photon flux generated
by undulator sources at synchrotron radiation facilities, typically by
more than five orders of magnitude.  The angular acceptance,
$1-10$~mrad, of the analyzer is much large than the angular
acceptance, $10-20~\mu$rad, of the monochromator; however, it is still
orders of magnitude smaller than the total solid angle of scattering
by the sample. It is also much smaller than the $10-100$-meV window
desired for the spectral analysis. All this results in very small
countrates $10 - 0.01$~Hz in IXS experiments \cite{Sinn01}.  Further
improvements to the 0.1-meV resolution using such an approach would
only result in yet another substantial reduction of the countrate and
time-consuming experiments.

A possible solution to this problem would be to create a spectrometer
that would not only feature the high spectral resolution, but would
also be capable of imaging x-ray spectra in a broad spectral
window. We will refer to such optical devices as x-ray spectrographs.

\CT\ type spectrographs \cite{CT30} are now standard in infrared,
visible, and ultraviolet spectroscopies \cite{SMD64, LTR10}. In its
classical arrangement, a spectrograph is comprised of four elements
[see Fig.~\ref{fig001}(a)]: (1) a collimating mirror, M$_{\indrm{C}}$,
that collects photons from a radiation source, S, and collimates the
photon beam; (2) a dispersing element, DE, such as a diffraction
grating or a prism, which scatters photons of different energies into
different directions $\theta^{\prime}(E)$ due to angular dispersion;
(3) a curved mirror, M$_{\indrm{F}}$, that focuses photons of
different energies into different locations $x(E)$ due to linear
dispersion; and (4) a spatially sensitive detector, Det, placed in the
focal plane to record the whole photon spectrum.

The feasibility of hard x-ray angular-dispersive spectrographs of the
\CT\ type has been discussed in \cite{Shv11,Shvydko12,SSM13}.  A hard
x-ray equivalent of the diffraction grating is a Bragg diffracting
crystal with diffracting atomic planes at an asymmetry angle $\eta\not
=0$ to the entrance crystal surface [see Fig.~\ref{fig001}(b)]
\cite{MK80-1,BSS95,Shvydko-SB}.  Angular dispersion rates attainable
in a single Bragg reflection are typically small, $\dirate \simeq
8~\mu$rad/meV \cite{SLK06,ShSS11}, and are the main obstacle to
realizing hard x-ray spectrographs.  The angular dispersion rate can
be enhanced dramatically, by almost two orders of magnitude, by
successive asymmetric Bragg reflections compared to that in a single
Bragg reflection \cite{SSM13}.  An enhanced angular dispersion rate in
multi-crystal arrangements is crucial for the feasibility of hard
x-ray angular-dispersive spectrographs.  An x-ray angular-dispersive
spectrograph was demonstrated experimentally in \cite{SSM13}, using
the so-called multi-crystal
collimation-dispersion-{wavelength-selection} (CDW) optic
\footnote{Abbreviation CDW is used to refer both, to all possible
  modification of the collimation-dispersion-wavelength-selection
  optic, in general (including its four-crystal modification CDDW) and
  to its original simplest three-crystal version, in particular.},
achieving spectral resolution of better than $100~\mu$eV with 9.1~keV
x-ray photons.  However, the spectral window in which the CDW optic
permitted imaging x-ray spectra was small, about $450~\mu$eV.
Increasing the spectral window of ultra-high-resolution x-ray
spectrographs, is extremely important.

In pursuing this goal, and in seeking solutions to this problem, a
theory of hard x-ray spectrographs is developed here. In
Section~\ref{rtm}, a ray-transfer matrix technique
\cite{KL66,MK80-1,MK80-2,Siegman,HoWe05,Smilgies08} is applied to
propagate x-rays through complex optical x-ray systems in the paraxial
approximation. The following systems are considered: successive Bragg
reflections from crystals (Section
~\ref{successive-Bragg-reflections}), focusing system
(Section~\ref{focusing-system}), focusing monochromators
(Section~\ref{focusing-monochromators}), and finally \CT -type
spectrographs (Section~\ref{spectrograph-section}). Solutions for
broadband hard x-ray imaging spectrographs are considered in
Section~\ref{broadband-spectrographs}.  Several ``diffraction
grating'' designs for hard x-ray spectrographs are proposed to ensure
a high energy resolution, broad spectral window of imaging, and large
angular acceptance.  Spectrographs with an energy resolution of
$\Delta E = 0.1$~meV and a spectral window of imaging up to $\Delta
E_{\ind{\cup}}= 45$~meV are shown to be feasible for IXS applications
in Section~\ref{cdw-uhrix} and Section~\ref{two-cristal-uhrix}. In
Section~\ref{two-cristal-rixs}, a spectrograph with a 1-meV spectral
resolution and 85-meV spectral imaging window is considered for Cu
K-edge resonant IXS (RIXS) applications.

\section{Ray-Transfer Matrices of X-ray Optical  Systems and Spectrographs}
\label{rtm}

The main goal of this article is to develop a theory of \CT -type hard
x-ray spectrographs. The conceptual optical scheme of the \CT -type
spectrographs is presented in Fig.~\ref{fig001}(a). In the hard x-ray
regime, the role of the diffraction grating is played by a single
crystal in asymmetric Bragg diffraction scattering geometry, as shown
in Fig.~\ref{fig001}(b), or by an arrangement of several single
crystals. One possible example of multi-crystal arrangements discussed
in \cite{SSM13}, although is not the only possibility, is shown in
Fig.~\ref{fig001}(c).  The purpose of the theory is to calculate the
spectral resolution and other performance characteristics of hard
x-ray spectrographs, and their dependence on physical parameters of
constituent optical elements.

In approaching the main goal, we consider optical systems starting
with simple ones, such as a focusing element and Bragg reflection from
a crystal, and proceed to more complex systems, such as successive
Bragg reflections from multiple crystals, focusing systems, focusing
monochromators, and finally spectrographs.

\subsection{Ray transfer matrix technique}

We will use a ray-transfer matrix technique
\cite{KL66,MK80-1,MK80-2,Siegman,HoWe05,Smilgies08} to propagate
paraxial x-rays through optical structures.  In a standard treatment,
a paraxial ray in a particular reference plane of an optical system
(the plane perpendicular to the optical axis $z$) is characterized by
its distance $x$ from the optical axis and by its angle or slope $\xi$
with respect to that axis. The ray is presented by a two-dimensional
vector $\vc{r}=(x,\xi)$. Interactions with optical elements are
described by $2\times 2$ dimensional $\{AB;CD\}$ matrices. The ray
vector $\vc{r}_{\ind{1}}=(x_{\ind{1}},\xi_{\ind{1}})$ at an input
reference plane (source plane) is transformed to
$\vc{r}_{\ind{2}}=\hat{O}\vc{r}_{\ind{1}}$ at the output reference
plane (image plane), where $\hat{O}$ is the ``ABCD'' matrix of an
element placed between the reference planes.

Angular dispersion in Bragg reflection from asymmetrically cut
crystals results in deviation of the beam from the unperturbed optical
axis due to a change, $\delta E$, in the photon energy from $E$ to
$E+\delta E$ \cite{MK80-1,BSS95,Shvydko-SB}. This causes
``misalignment'' of the paraxial optical system, which can be
conveniently described by a $3\times 3$ $\{ABG;CDF; 001\}$ matrix by
adding additional coordinate $\delta E$ to vector
$\vc{r}=(x,\xi,\delta E)$
\cite{MK80-1,MK80-2,Siegman,Martinez88,Smilgies08}.

Table~\ref{tab2} presents ray-transfer matrices used in this paper.
In the first three rows, 1--3, matrices are given for simple elements
of the spectrograph, such as propagation in free space, thin lens or
focusing mirror, and Bragg reflection from a crystal. In the following
rows ray-transfer matrices are shown for arrangements composed of
several optical elements, such as successive multiple Bragg
reflections from several crystals, rows 4--5; focusing system, row 6;
focusing monochromators, rows 7--8; and finally spectrographs, row 9,
on which the paper is focused. The matrices of the multi-element
systems are obtained by successive multiplication of the matrices of
the constituent optical elements.

\subsection{Bragg reflection and reference system}

\begin{figure}
\setlength{\unitlength}{\textwidth}
\includegraphics[width=0.5\textwidth]{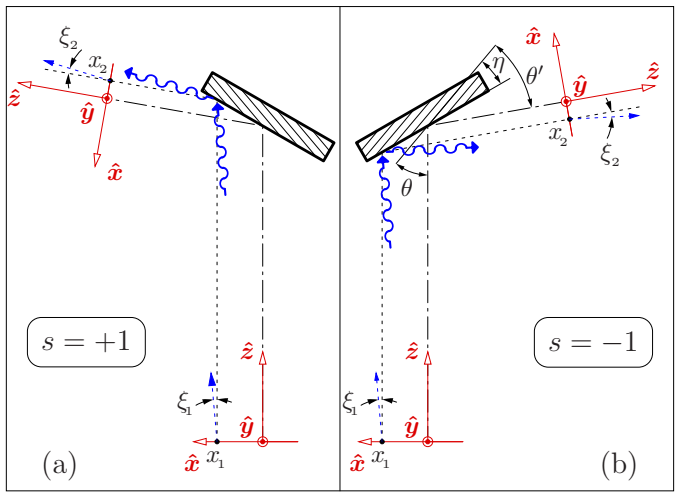}
\caption{(Color online) Schematic elucidating the definition of an
  absolute right-handed coordinate system $\{x,y,z\}$ with the
  $\hat{\vc{z}}$-axis always looking in the direction of the optical
  axis (dash-dotted line) both before and after each optical element,
  a Bragg reflecting crystal in this particular case. By definition,
  positive is the counterclockwise sense of angular variations $\xi$
  of ray slopes in the $(x,z)$ plane. Shown are examples of an optical
  element ``Bragg reflection from a crystal'' with (a)
  counterclockwise deflection, for which deflection sign $s=+1$; and
  (b) clockwise deflection of the reflected beam with $s=-1$. In the
  input reference system, the incident x-ray beam (wavy vector line)
  impinges onto the crystal at a glancing angle of incidence
  $\theta-s\xi_{\ind{1}}$ to the Bragg reflecting planes and with
  coordinate $x_{\ind{1}}$. It is reflected at a glancing angle of
  reflection $\theta^{\prime}+s\xi_{\ind{2}}$ with coordinate
  $x_{\ind{2}}$, as seen in the output reference system. Both $x$ and
  $\xi$ change signs upon reflection. }
\label{fig011}
\end{figure}

All the ray-transfer matrices are presented in the right-handed
coordinate system $\{x,y,z\}$ with the $\hat{\vc{z}}$-axis looking in
the direction of the optical axis both before and after each optical
element, as illustrated in Fig.~\ref{fig011} on an example of a Bragg
reflecting crystal.  This absolute reference system is retained
through all interactions with all optical elements. We use the
convention that positive is the counterclockwise sense of angular
variations $\xi$ of the ray slope in the $(x,z)$ plane. For Bragg
reflections, $\xi_{\ind{1}}$ is understood as a small angular
deviation from a nominal glancing angle of incidence $\theta$ to the
reflecting atomic planes of the crystal; $\xi_{\ind{2}}$ is understood
as a small angular deviation from the nominal glancing angle of
reflection $\theta^{\prime}$. The angles $\theta$ and
$\theta^{\prime}$ define the optical axis. The angle $\theta$ is
determined by Bragg's law $2K\sin\theta=H$, while $\theta^{\prime}$ is
determined by the relationship \cite{KB76}
\begin{equation}                                                     
  \cos(\theta^{\prime}-\eta) =  \cos(\theta+\eta) + \frac{H}{K}\sin{\eta}.                
\label{ad000}                                                        
\end{equation}                                                       
Equation~\eqref{ad000} is a consequence, first, of the conservation of
the tangential components
$(\vc{K}_{\ind{H}})_t=(\vc{K}_{\ind{0}}+\vc{H})_t$ with respect to the
entrance crystal surface for the momentum, $\vc{K}_{\ind{0}}$, of the
incident x-ray photon and the momentum, $\vc{K}_{\ind{H}}$, of the
photon Bragg reflected from the crystal with a diffraction vector
$\vc{H}$ [see Fig.~\ref{fig001}(b)]. It is also a consequence of the
conservation of the photon energies $|\vc{K}_{\ind{H}}|\hbar
c=|\vc{K}_{\ind{0}}|\hbar c=K\hbar c= E$.  The reflecting atomic
planes are at an asymmetry angle, $\eta$, to the entrance crystal
surface. The asymmetry angle, $\eta$, is defined here to be positive
in the geometry shown in Figs.~\ref{fig011}(a) and \ref{fig011}(b),
and negative in the geometry with reversed incident and reflected
x-rays (not shown).

\begin{table*}[t!]
\centering
\begin{tabular}{|l|l|l|l|}
  \hline 
& & & \\
  Optical system & Matrix notation & Ray-transfer matrix  & Definitions and Remarks\\[-5pt]    
& & & \\
  \hline  \hline
& & & \\[-3.8mm]  
\parbox[c]{0.25\textwidth}{Free space \cite{KL66,HoWe05,Siegman}\\  \includegraphics[width=0.25\textwidth]{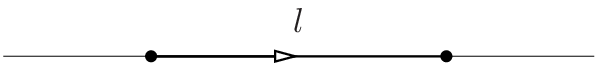}\\[-8mm] \hspace*{-40mm} (1)\\[3mm] }       &  $\hat{\fspace}(l)$ & $\left( \begin{array}{ccc} 1 & l  & 0 \\ 0 & 1 & 0 \\ 0 & 0 & 1  \end{array} \right)$      & \parbox[c]{0.21\textwidth}{$l$ -- distance}     \\[-0.5mm] 
\hline 
& & & \\[-3.8mm]  
\parbox[c]{0.25\textwidth}{Thin lens \cite{KL66,HoWe05,Siegman}\\ \includegraphics[width=0.25\textwidth]{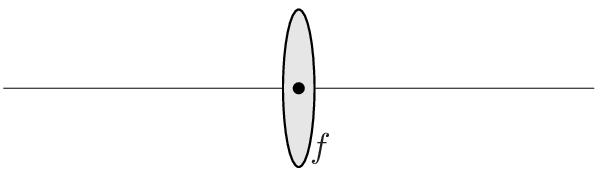}\\[-8mm] \hspace*{-40mm} (2)\\[3mm] }       &  $\hat{\thinlens}(f)$ & $\left( \begin{array}{ccc} 1 & 0 & 0\\ -\frac{1}{f} & 1 & 0  \\ 0 & 0 & 1 \end{array} \right)$      &  \parbox[c]{0.21\textwidth}{$f$ -- focal length}      \\ [-0.5mm]  
\hline 
& & & \\[-4.2mm]  
\parbox[c]{0.25\textwidth}{Bragg reflection from a crystal\\ \cite{MK80-1,MK80-2}\\ \includegraphics[width=0.25\textwidth]{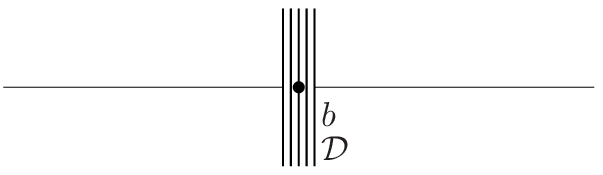} \\[-8mm] \hspace*{-40mm} (3)\\[3mm] }  &  $\hat{\crystal}(b,\sgn \dirate)$ & $\left( \begin{array}{ccc} {1}/{b} & 0 & 0  \\0  & b & \sgn\dirate \\ 0 & 0 & 1 \end{array} \right)$  &  \parbox[c]{0.21\textwidth}{$b=-\frac{\sin(\theta+\eta)}{\sin(\theta-\eta)}$\\ asymmetry factor;\\ $\dirate = -(1/E)(1+b)\tan\theta $\\ angular dispersion rate;\\ $\sgn = -1$ for clockwise, and  $\sgn = +1$ counterclockwise ray deflection.}     \\[+0.5mm] 
\hline 
& & & \\[-3.8mm]  
\parbox[c]{0.25\textwidth}{Successive Bragg reflections\\ $\hat{\crystal}(\!b_{\ind{n}},\sgn_{\ind{n}}\dirate_{\ind{n}}\!)\dotsb \hat{\crystal}(\!b_{\ind{1}},\sgn_{\ind{1}}\dirate_{\ind{1}}\!)$ \\  \includegraphics[width=0.25\textwidth]{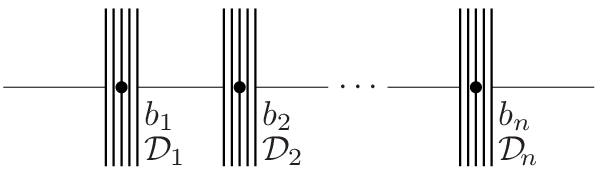} \\[-8mm] \hspace*{-40mm} (4)\\[3mm] }  &  $\hat{\crystal}(\dcomm{n},\fcomm{n})$ & $\left( \begin{array}{ccc} \acomm{n} & 0 & 0  \\ 0  & \dcomm{n}  & \fcomm{n} \\ 0 & 0 & 1 \end{array} \right)$  &  \parbox[c]{0.21\textwidth}{$\dcomm{n}=b_{\ind{1}}b_{\ind{2}}b_{\ind{3}} \dotsc b_{\ind{n}}$ \\ $\fcomm{n}=b_{\ind{n}}\fcomm{n-1} + \sgn_{\ind{n}}\dirate_{\ind{n}}$\\ $\sgn_i=\pm 1$, $i=1,2,...,n$ }     \\[-0.5mm] 
\hline 
& & & \\[-3.8mm]  
\parbox[c]{0.25\textwidth}{Successive Bragg reflections with space between crystals\\ $\hat{\crystal}(\!b_{\ind{n}},\sgn_{\ind{n}}\dirate_{\ind{n}}\!)\dotsb \hat{\fspace}(\!l_{\ind{12}}\!)\hat{\crystal}(\!b_{\ind{1}},\sgn_{\ind{1}}\dirate_{\ind{1}}\!)$ \\  \includegraphics[width=0.25\textwidth]{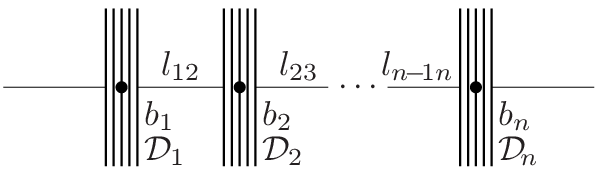}\\[-8mm] \hspace*{-40mm} (5)\\[3mm] }  &  $\hat{\crystalsp}(\dcomm{n},\fcomm{n},l)$ & $\left( \begin{array}{ccc} \acomm{n} & \bcomm{n} & \gcomm{n}  \\ 0  & \dcomm{n}  & \fcomm{n} \\ 0 & 0 & 1 \end{array} \right)$  &  \parbox[c]{0.21\textwidth}{$\bcomm{n}\!=\!\frac{\bcomm{n-1}+\dcomm{n-1}l_{\ind{n-1 n}}}{b_{\ind{n}}}$\\[1mm]
$\gcomm{n}\!=\!
\frac{\gcomm{n-1}+\fcomm{n-1}l_{\ind{n-1 n}}}{b_{\ind{n}}}$\\[1mm]  $\bcomm{1}\!=\!0, \hspace{0.5cm} \gcomm{1}\!=\!0 $}     \\[-0.5mm] 
\hline 
& & & \\[-3.8mm]
\parbox[c]{0.25\textwidth}{Focusing system\\ $\hat{\fspace}(l_{\ind{2}})\hat{\thinlens}(f)\hat{\fspace}(l_{\ind{1}})$ \\ \includegraphics[width=0.25\textwidth]{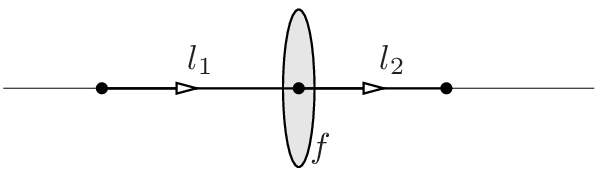} \\[-8mm] \hspace*{-40mm} (6)\\[3mm] }  &  $\hat{\focus}(l_{\ind{2}},f,l_{\ind{1}})$ & $\left( \begin{array}{ccc} 1-\frac{l_{\ind{2}}}{f} & B_{\indrm{F}} & 0  \\ -\frac{1}{f}  & 1-\frac{l_{\ind{1}}}{f} & 0\\ 0 & 0 & 1  \end{array} \right)$  &  \parbox[c]{0.21\textwidth}{$B_{\indrm{F}}=l_{\ind{1}}l_{\ind{2}}\left(\frac{1}{l_{\ind{1}}}+\frac{1}{l_{\ind{2}}}-\frac{1}{f}\right)$ }     \\[-0.5mm] 
\hline 
& & & \\[-3.8mm]  
\parbox[c]{0.25\textwidth}{Focusing~monochromator~I \cite{KCR09}\\ $\hat{\fspace}(l_{\ind{3}})\hat{\crystal}(\dcomm{n}\!,\fcomm{n}\!)\hat{\focus}(l_{\ind{2}},f,l_{\ind{1}})$ \\ \includegraphics[width=0.25\textwidth]{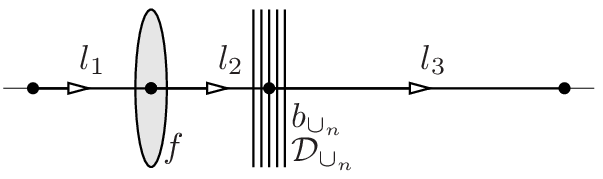} \\[-8mm] \hspace*{-40mm} (7)\\[3mm] }  &  $\hat{\focusimgmono_{\indrm{I}}}(l_{\ind{3}},\dcomm{n}\!,\fcomm{n}\!,l_{\ind{2}},f,l_{\ind{1}})$ & $\left(\!\! \begin{array}{ccc} \frac{1}{\dcomm{n}}\!\left(\!1\!-\!\frac{l_{\ind{23}}}{f}\!\right) & B_{\indrm{\fmo}}  & l_{\ind{3}}\fcomm{n}  \\ -\frac{\dcomm{n}}{f}  & \dcomm{n}\!\left(\!1\!-\!\frac{l_{\ind{1}}}{f}\!\right) & \fcomm{n} \\ 0 & 0 & 1  \end{array}\!\! \right) $  &  \parbox[c]{0.21\textwidth}{$B_{\indrm{\fmo}}= \frac{l_{\ind{1}}l_{\ind{23}}}{\dcomm{n}}\!\left(\!\frac{1}{l_{\ind{1}}}\!+\!\frac{1}{l_{\ind{23}}}\!-\!\frac{1}{f}\!\right)$\\
$l_{\ind{23}}=l_{\ind{2}}+\dcomm{n}^2 l_{\ind{3}}$  }     \\[-0.5mm] 
\hline 
& & & \\[-3.8mm]
\parbox[c]{0.25\textwidth}{Focusing~monochromator~II\\ $\hat{\focus}(l_{\ind{3}},f,l_{\ind{2}}) \hat{\crystal}(\dcomm{n}\!,\fcomm{n}\!)\hat{\fspace}(l_{\ind{1}})$ \\ \includegraphics[width=0.25\textwidth]{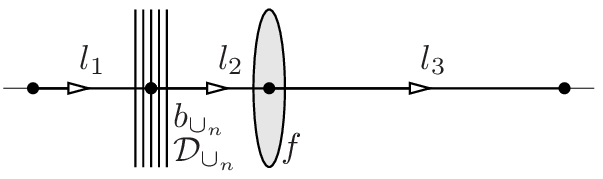} \\[-8mm] \hspace*{-40mm} (8)\\[3mm] }  &  $\hat{\focusimgmono_{\indrm{II}}}(l_{\ind{3}},f,l_{\ind{2}},\dcomm{n}\!,\fcomm{n}\!,l_{\ind{1}})$ & $\left(\!\! \begin{array}{ccc} \frac{1}{\dcomm{n}\!}\!\left(\!1\!-\!\frac{l_{\ind{3}}}{f}\!\right) & B_{\indrm{\fmt}}  &  X \fcomm{n}  \\ 
-\frac{1}{f\dcomm{n}}  & \dcomm{n}\!\!\left(\!1\!-\!\frac{l_{12}}{f}\!\right) & \left(\!1\!-\!\frac{l_{\ind{2}}}{f}\!\right)\!\fcomm{n} \\ 
0 & 0 & 1  \end{array}\!\! \right)
 $  &  \parbox[c]{0.21\textwidth}{$B_{\indrm{\fmt}}\! =  \!\dcomm{n}\!l_{\ind{12}} l_{\ind{3}}\! \left(\!\frac{1}{l_{\ind{12}}}\!+\!\frac{1}{l_{\ind{3}}}\!-\!\frac{1}{f}\!\right)$ \\[1mm]
$l_{\ind{12}} =l_{\ind{1}} /\dcomm{n}^2 + l_{\ind{2}}  $ \\[1mm] $X=l_{\ind{2}} +l_{\ind{3}} -l_{\ind{2}} l_{\ind{3}}/f$\\[1mm]  $B_{\indrm{\fmt}}\!=\!0 \Rightarrow X\!=\!l_{\ind{3}} l_{\ind{1}}\!/(\!\dcomm{n}^2\! l_{\ind{12}}\!) $  }     \\[-0.5mm] 
\hline 
& & & \\[-3.8mm]
\parbox[c]{0.25\textwidth}{Spectrograph\\ $\hat{\focus}(\!f_{\ind{2}}\!,\!f_{\ind{2}}\!,\!l_{\ind{2}}\!) \hat{\crystal}(\!\dcomm{n}\!,\!\fcomm{n}\!) \hat{\focus}(\!l_{\ind{1}}\!,\!f_{\ind{1}}\!,\!f_{\ind{1}}\!) $\\ \includegraphics[width=0.25\textwidth]{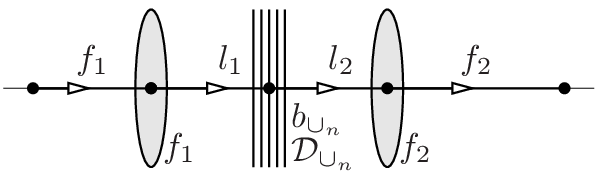} \\[-8mm] \hspace*{-40mm} (9)\\[3mm] }   &  $\hat{\spectro}(f_{\ind{2}},l_{\ind{2}},\dcomm{n}\!,\fcomm{n}\!,l_{\ind{1}},f_{\ind{1}})$ & $ 
\left(\!\! \begin{array}{ccc} -\frac{\dcomm{n}f_{\ind{2}}\!}{f_{\ind{1}}\!}  & 0 &  f_{\ind{2}} \fcomm{n}  \\ \frac{(\!l_{\ind{1}}\!-\!f_{\ind{1}}\!)\!+\!(\!l_{\ind{2}}\!-\!f_{\ind{2}}\!)\dcomm{n}^2\!}{\!\dcomm{n}f_{\ind{1}}f_{\ind{2}}\!}  & -\frac{f_{\ind{1}}}{\dcomm{n}\!f_{\ind{2}}} & \left(\!1\!-\!\frac{l_{\ind{2}}}{f_{\ind{2}}}\!\right)\!\fcomm{n}\! \\ 0 & 0 & 1 \end{array}\!\! \right)
 $  &  \parbox[c]{0.21\textwidth}{ }     \\ 
  \hline  
\end{tabular}
\caption{Table of ray-transfer matrices $\{ABG,CDF,001\}$ used in the
  paper. The matrices for the focusing monochromators and the imaging
  spectrograph presented here are calculated with multi-crystal matrix
  $\hat{\crystal}(\dcomm{n},\fcomm{n})$ from row 4, assuming zero free space
  between crystals in successive Bragg reflections.  Generalization to
  a more realistic case of nonzero distances between the crystals
  requires application of matrix
  $\hat{\crystalsp}(\dcomm{n},\fcomm{n},l)$ from row 5. The results are
  discussed in the text.}
\label{tab2}
\end{table*}

For the Bragg-reflection matrix $\{ABG,CDF,001\}$ 
the nonzero elements 
$A=1/D$, $D=\xi_{\ind{2}}/\xi_{\ind{1}}$, $F=\xi_{\ind{2}}/\delta E$ are calculated from Eq.~\eqref{ad000}  as follows
:
\begin{equation}
D\, =\,b, \hspace{0.5cm} A\, =\,1/b, \hspace{0.5cm} b\,=\, -\frac{\sin(\theta+\eta)}{\sin(\theta-\eta)},
\label{ad090}
\end{equation}                                                                                  
\begin{equation}                                                                                                                    
F\,=\, s\, \dirate, \hspace{0.5cm} \dirate= \frac{2\sin\theta \sin{\eta}}{E\sin(\theta-\eta)} \equiv -\frac{1}{E} (1+b)\tan\theta,        
\label{ad010}
\end{equation}
by using the following variations: $\theta \rightarrow
\theta-s\xi_{\ind{1}}$, $\theta^{\prime}\rightarrow
\theta^{\prime}+s\xi_{\ind{2}}$, $E \rightarrow E+\delta E$ (see
caption to Fig.~\ref{fig011}).  The angular dispersion rate $\dirate$
in Eq.~\eqref{ad010} describes how the photon energy variation $\delta
E$ changes the reflection angle at a fixed incidence angle.  The
deflection sign factor $s$ allows for the appropriate sign, depending
on the scattering geometry. It is defined to be $s=+1$ if the Bragg
reflection deflects the ray counterclockwise
[Fig.~\ref{fig011}(a)]. It is $s=-1$ if the reflected ray is deflected
clockwise [Fig.~\ref{fig011}(b)].  Asymmetry factor $b$ in
Eq.~\eqref{ad090} describes, in particular, how the beam size and beam
divergence change upon Bragg reflection.

The ray transfer matrix for a Bragg reflection from a crystal
presented in row 3 of Table~\ref{tab2} is equivalent to that
introduced by Matsushita and Kaminaga \cite{MK80-1,MK80-2}, with the
exception for different signs of the elements and the additional
deflection sign factor $s$.  Positive absolute values $|b|$ were used
in local coordinate systems in \cite{MK80-1,MK80-2}. Here we use the
absolute coordinate system to correctly describe transformations in
multi-element optical arrangements.  The choice of the absolute
coordinate system is especially important to allow for inversion of
the transverse coordinate and inversion of the slope when an optical
ray is specularly reflected from a mirror or Bragg reflected from a
crystal. Because of such inversion $x_{\ind{1}}$ and $x_{\ind{2}}$ as
well as $\xi_{\ind{1}}$ and $\xi_{\ind{2}}$ have opposite signs, as
shown in Figs.~\ref{fig011}(a) and \ref{fig011}(b).  A negative value
of the asymmetry factor, $b$, in the Bragg reflection ray transfer
matrix reflects this inversion upon each Bragg reflection.

The Bragg diffraction $\hat{\crystal}(b,\dirate)$ matrix is similar to
the ray-transfer matrix of the diffraction grating (see, e.g.,
\cite{Martinez88}).  The similarity is because both the asymmetry
factor, $b$, and the angular dispersion rate, $\dirate$, are derived
from Eq.~\eqref{ad000}, which coinsides with the well-know in optics
grating equation. The magnification factor, $m$, used in the
diffraction grating matrix is equivalent to $1/|b|$.

\subsection{Thin lens or elliptical mirror}

The ray-transfer matrix of a thin lens $\hat{\thinlens}(f)$
\cite{KL66,HoWe05,Siegman} has a focal distance, $f$, as a
parameter. Compound refractive lenses  \cite{SKSL} can be used
for focusing and collimation in the hard x-ray regime, and described
by such a matrix to a certain approximation. Alternatively,
ellipsoidal total reflection mirrors could be applied, which transform
radiation from a point source at one focal point to a point source
located at the second focal point. The ray-transfer matrix of an
ellipsoidal mirror has a structure identical to the ray-transfer
matrix of a thin lens; however, $1/f = 1/R_{\ind{1}} + 1/R_{\ind{2}}$,
where $R_{\ind{1}}$ and $R_{\ind{2}}$ are the distances from the
center of the section of the ellipsoid employed to the foci of the
generating ellipse \cite{Goldsmith98}.

The basic ray matrices given in the first three rows of
Table~\ref{tab2} can be combined to represent systems that are more
complex.

\subsection{Successive Bragg reflections}
\label{successive-Bragg-reflections} 

The ray-transfer matrix
$\hat{\crystal}_{\cup_{\ind{n}}}(\dcomm{n},\fcomm{n})$ describing
successive Bragg reflections from $n$ different crystals has a
structure identical to that of the single Bragg reflection
ray-transfer matrix $\hat{\crystal}(b,\dirate)$; however, the
asymmetry factor, $b$, and the angular dispersion rate, $\dirate$, are
substituted by the appropriate cumulative values $\dcomm{n}$, and
$\fcomm{n}$, respectively, as defined in row 4 of
Table~\ref{tab2}. The cumulative angular dispersion rate, $\fcomm{n}$,
derived in the present paper coincides with the expression first
derived in \cite{SSM13} using an alternative approach. It should be
noted that the ray-transfer matrix
$\hat{\crystal}_{\cup_n}(\dcomm{n},\fcomm{n})$ presented in
Table~\ref{tab2}, row 4, was derived neglecting propagation through
free space between the crystals.

With nonzero distances $l_{\ind{i-1\ i}}$ between the crystals $i-1$
and $i$ ($i=2,3,...,n$) taken into account, the ray-transfer matrix of
successive Bragg reflections changes to
$\hat{\crystalsp}_{\cup_n}(\dcomm{n},\fcomm{n},l)$, as presented in
row 5 of Table~\ref{tab2}.  Most of the elements of the modified
ray-transfer matrix still remain unchanged, except for elements
$\bcomm{n}$ and $\gcomm{n}$, which become nonzero. These elements are
defined by recurrence relations in the table.  Nonzero distances
$l_{\ind{i-1\ i}}$ between the crystals result in an additional change
$\bcomm{n}\xi$ of the linear size of the source image due to an
angular spread $\xi$, and in an spatial transverse shift
$\gcomm{n}\delta E$ of the image (linear dispersion) due to a spectral
variation $\delta E$.

\subsection{Focusing system}
\label{focusing-system}
 
In the focusing system [see graph in row 6 of Table~\ref{tab2}] a
source in a reference source plane at a distance $l_{\ind{1}}$
downstream a lens or an elliptical mirror is imaged onto the reference
image plane at a distance $l_{\ind{2}}$ upstream of the lens. The
ray-transfer matrix of the focusing system is a product of the
ray-transfer matrices of the free space $\hat{\fspace}(l_{\ind{1}})$,
the thin lens $\hat{\thinlens}(f)$, and another free space matrix
$\hat{\fspace}(l_{\ind{2}})$. If defined in Table~\ref{tab2} for the
focusing system parameter $B_{\indrm{F}}=0$, the classical lens
equation is valid:
\begin{equation}
\frac{1}{l_{\ind{1}}}+\frac{1}{l_{\ind{2}}}=\frac{1}{f}.
\label{fs010}
\end{equation}
In this case, the system images the source with inversion and a
magnification factor $1-l_{\ind{2}}/f=-l_{\ind{2}}/l_{\ind{1}}$
independent of the angular spread of rays in the source plane.

\subsection{Focusing monochromators}
\label{focusing-monochromators}

Rows 7--8 in Table~\ref{tab2} present ray-transfer matrices of
focusing monochromators, optical systems comprising a lens or an
elliptical mirror, and an arrangement of crystals, respectively.

We will distinguish between two different types of focusing
monochromators. If the lens is placed upstream of the crystal
arrangement, we will refer to such optic as a focusing monochromator,
I, presented in row 7 of Table~\ref{tab2}. If the lens is placed
downstream, this optic will be referred to as focusing monochromator,
II, presented in row 8.

\subsubsection{Focusing monochromator I}

The focusing monochromator I with a single crystal was introduced in
\cite{KCR09}, and its performance was analyzed using the wave theory
developed there. The ray-transfer matrix approach used in the present
paper leads to similar results, except for diffraction effects being
neglected here.

We consider here a general case with a multi-crystal arrangement.  The
ray-transfer matrix presented in Table~\ref{tab2} was derived
neglecting propagation through free space between the crystals.  The
following expressions are valid for the elements of the ray-transfer
matrix of the focusing monochromator~I if nonzero distances between
the crystals of the monochromator are taken into account:
\begin{equation}
\tilde{A}_{\indrm{\fmo}} = \frac{1}{\dcomm{n}}\left(1- \frac{\tilde{l}_{\ind{23}}}{f} \right), 
\label{fm101}
\end{equation}
\begin{equation}
\tilde{l}_{\ind{23}} = l_{\ind{23}} + \dcomm{n} \bcomm{n},\hspace{0.5cm} l_{\ind{23}} = l_{\ind{2}} + l_{\ind{3}} \dcomm{n}^2,
\label{fm102}
\end{equation}
\begin{equation}
\tilde{B}_{\indrm{\fmo}}= \frac{l_{\ind{1}}\tilde{l}_{\ind{23}}}{\dcomm{n}} \left(\frac{1}{l_{\ind{1}}}+\frac{1}{\tilde{l}_{\ind{23}}}-\frac{1}{f} \right),
\label{fm103}
\end{equation}
\begin{equation}
\tilde{G}_{\indrm{\fmo}}= G_{\indrm{\fmo}} + \gcomm{n}, \hspace{1cm} G_{\indrm{\fmo}}=\fcomm{n} l_{\ind{3}},  
\label{fm104}
\end{equation}
\begin{equation}
\tilde{C}_{\indrm{\fmo}}= C_{\indrm{\fmo}}, \hspace{0.5cm} \tilde{D}_{\indrm{\fmo}}= D_{\indrm{\fmo}}, \hspace{0.5cm} \tilde{F}_{\indrm{\fmo}}= F_{\indrm{\fmo}}. 
\label{fm105}
\end{equation}
The main difference is  that the parameter $l_{\ind{23}}$ has to be
substituted by $\tilde{l}_{\ind{23}}=l_{\ind{23}} + \dcomm{n}
\bcomm{n}$. The nonzero distances
between the crystals also change the linear dispersion rate from
$G_{\indrm{\fmo}}=\fcomm{n} l_{\ind{3}}$ to
$\tilde{G}_{\indrm{\fmo}}$.

If the focusing condition $B_{\indrm{\fmo}}=0$ is fulfilled (assuming
the system with zero free space between crystals), the following
relationship is valid for the focal $f$ and other distances involved
in the problem:
\begin{equation}
\frac{1}{l_{\ind{1}}}+\frac{1}{l_{\ind{23}}}=\frac{1}{f}.
\label{fm100}
\end{equation}
Without the crystals, the image plan would be at a distance
$l_{\ind{2}} + l_{\ind{3}}$ from the lens, in agreement with
Eq.~\eqref{fs010}.  The presence of the crystal changes the position
of the image plane to $l_{\ind{23}} = l_{\ind{2}} + l_{\ind{3}}
\dcomm{n}^2$. Such behavior for the focusing monochromator-I system
was predicted in \cite{KCR09}; it is related to the ability of
asymmetrically cut crystals to change the beam angular divergence and
linear size and thus the virtual position of the source \cite{SDSS99}.

If the  focusing condition  $B_{\indrm{\fmo}}=0$ is fulfilled,
Eq.~\eqref{fm100} is valid and,  as a consequence, the focusing
monochromator~I images a source spot of size $\Delta x$ into a spot of
size
\begin{equation}
  \Delta x^{\prime} = -\frac{1}{\dcomm{n}} \frac{l_{\ind{23}}}{l_{\ind{1}}} \Delta x,
\label{fm110}
\end{equation}
for each monochromatic component $E$.  If the source is not
monochromatic, its image by photons with energy $E+\delta E$ is
shifted transversely as a result of linear dispersion, by
\begin{equation}
\delta x^{\prime} = l_{\ind{3}} \fcomm{n} \delta E 
\label{fm120}
\end{equation}
from the source image position produced by photons of energy $E$. 

The  monochromator spectral resolution $\Delta E$ can be
determined from the condition that the monochromatic source image size
$\Delta x^{\prime}$ [Eq.~\eqref{fm110}], is equal to the source image
shift $\delta x^{\prime}$ [Eq.~\eqref{fm120}]:
\begin{equation}
\Delta E =  \frac{1}{\fcomm{n} |\dcomm{n}|} \frac{\Delta x\, l_{\ind{23}}}{l_{\ind{3}} l_{\ind{1}}}. 
\label{fm130}
\end{equation}
Here and in the rest of the paper it is assumed that the source image
size $\Delta x^{\prime}$ can be resolved by the position-sensitive
detector. In a particular case of $l_{\ind{2}} \ll\ l_{\ind{3}}
\dcomm{n}^2$, $l_{\ind{23}}$ can be approximated by $l_{\ind{23}} =
l_{\ind{3}} \dcomm{n}^2$. As a result, the expression for the energy
resolution can be simplified to
\begin{equation}
\Delta E =  \frac{|\dcomm{n}|}{\fcomm{n} } \frac{\Delta x}{l_{\ind{1}}}. 
\label{fm140}
\end{equation}
A large dispersion cumulative rate $\fcomm{n}$, a small cumulative
asymmetry factor $|\dcomm{n}|$, a large distance $l_{\ind{1}}$ from
the source to the lens, and a small source size $\Delta x$ are
advantageous for better spectral resolution. This result is in
agreement with the wave theory prediction \cite{KCR09}, generalized to
a multi-crystal monochromator system. All these results can be further
generalized in a straightforward manner to account for nonzero spaces
between the crystals, using Eqs.~\eqref{fm101}--\eqref{fm105}.

\subsubsection{Focusing monochromator  II}

In the focusing monochromator-II system the focusing element is placed
downstream of the crystals system [see graph in row 8 of
Table~\ref{tab2}]. The ray-transfer matrix presented in
Table~\ref{tab2} is derived neglecting propagation through free space
between the crystals.  The following expressions are valid for the
elements of the ray-transfer matrix if nonzero distances between the
crystals of the monochromator are taken into account:
\begin{equation}
\tilde{A}_{\indrm{\fmt}}= A_{\indrm{\fmt}}, \hspace{0.5cm} \tilde{C}_{\indrm{\fmt}}= C_{\indrm{\fmt}}. 
\label{fm201}
\end{equation}
\begin{equation}
\tilde{B}_{\indrm{\fmt}}=  \dcomm{n} \tilde{l}_{\ind{12}} l_{\ind{3}} \left(\frac{1}{\tilde{l}_{\ind{12}}}+\frac{1}{l_{\ind{3}}}-\frac{1}{f} \right),
\label{fm202}
\end{equation}
\begin{equation}
\tilde{l}_{\ind{12}} = l_{\ind{12}} +  \bcomm{n}/\dcomm{n} , 
\hspace{0.25cm} l_{\ind{12}} = l_{\ind{1}} / \dcomm{n}^2 + l_{\ind{2}}, 
\label{fm203}
\end{equation}
\begin{equation}
\tilde{D}_{\indrm{\fmt}}=  \dcomm{n} \left(1-\frac{\tilde{l}_{\ind{12}} l_{\ind{3}}}{f} \right),
\label{fm204}
\end{equation}
\begin{equation}
\tilde{G}_{\indrm{\fmt}}= G_{\indrm{\fmt}} + \gcomm{n}\left(1-\frac{l_{\ind{3}}}{f} \right), \hspace{1cm} G_{\indrm{\fmt}}=\fcomm{n} X,  
\label{fm205}
\end{equation}
\begin{equation}
\tilde{F}_{\indrm{\fmt}}= F_{\indrm{\fmt}} - \frac{\gcomm{n}}{f}, \hspace{1cm} F_{\indrm{\fmt}}=\fcomm{n} \left(1-\frac{l_{\ind{2}}}{f}\right).  
\label{fm206}
\end{equation}
Elements $\tilde{A}_{\indrm{\fmt}}$, $\tilde{B}_{\indrm{\fmt}}$,
$\tilde{C}_{\indrm{\fmt}}$, and $\tilde{D}_{\indrm{\fmt}}$ have the
same form as in Table~\ref{tab2}, but with the distance parameter
$l_{\ind{12}}$ replaced by $\tilde{l}_{\ind{12}}$.  Elements
$\tilde{G}_{\indrm{\fmt}}$ $\tilde{F}_{\indrm{\fmt}}$ obtain
additional correction terms.

If the focusing condition $B_{\indrm{\fmt}}=0$ is fulfilled (we further
assume an idealized case of a system with zero free space between
crystals), the following relationship is valid
for the focal $f$ and other distances involved in the problem:
\begin{equation}
  \frac{1}{l_{\ind{12}}}+\frac{1}{l_{\ind{3}}}=\frac{1}{f}.
\label{fm200}
\end{equation}
Without the crystals, the source should be at a distance $l_{\ind{1}}
+ l_{\ind{2}}$ upstream of the lens to achieve focusing at a distance
$l_{\ind{3}}$ of downstream the lens, in agreement with
Eq.~\eqref{fs010}.  The presence of the crystals changes the virtual
position of the source plane, which will now be located at a distance
$l_{\ind{12}} = l_{\ind{1}}/\dcomm{n}^2 + l_{\ind{2}}$ from the lens
\footnote{Particular optical schemes similar to the considered here
  focusing monochromator-II have been studied in \cite{HMH12} using
  geometric ray tracing. In agreement with our result, the virtual
  source was determined to be at a distance $l_{\ind{1}}/\dcomm{n}^2$
  from the crystal, using the notations of the present paper.}.
Therefore, unlike the monochromator-I case, in which the crystals
change the virtual image plane position, the crystals in the
monochromator-II system change the virtual source plane position.

Using a process similar to that used to derive these values for the
monochromator-I system, we obtain the following expressions for the
image size $\Delta x^{\prime}$, the transverse image shift $\delta
x^{\prime}$ (linear dispersion), and for the spectral resolution
$\Delta E$ of the monochromator-II system:
\begin{equation}
  \Delta x^{\prime} = -\frac{1}{\dcomm{n}} \frac{l_{\ind{3}}}{l_{\ind{12}}} \Delta x,
\label{fm210}
\end{equation}
\begin{equation}
\delta x^{\prime} = X \fcomm{n} \delta E, \hspace{0.5cm} X=\frac{l_{\ind{3}} l_{\ind{1}}}{\dcomm{n}^2 l_{\ind{12}}},   
\label{fm220}
\end{equation}
\begin{equation}
\Delta E =  \frac{|\dcomm{n}|}{\fcomm{n} } \frac{\Delta x}{l_{\ind{1}}}. 
\label{fm230}
\end{equation}
Interestingly, the expression for the energy resolution of the
monochromator-II system [Eq.~\eqref{fm230}] is equivalent to that of
the monochromator-I system given by Eq.~\eqref{fm140}. We recall,
however, that Eq.~\eqref{fm140} was derived for a particular case of
$l_{\ind{2}} \ll\ l_{\ind{3}} \dcomm{n}^2$, while Eq.~\eqref{fm230} is
valid in general case.

We would like to emphasize one particular interesting case. If
$l_{\ind{1}}=0$ (i.e., the source position coincides with the position
of the crystal system), then $X=0$, what results in zero linear
dispersion rate $X \fcomm{n} =0$. This property can be used to
suppress linear dispersion, if it is undesirable.  It often happens
when a crystal monochromator is combined with a focusing system.  This
conclusion is strictly valid, provided nonzero distances between the
crystals of the monochromator are neglected.

The results derived above, can be further generalized to take the
nonzero spaces between the crystals into account by applying
Eqs.~\eqref{fm201}--\eqref{fm206}.

\subsection{Spectrograph}
\label{spectrograph-section}

In this section we consider spectrographs in a \CT\ configuration with
the optical scheme shown in Fig.~\ref{fig001}, or alternatively in the
graph in Table~\ref{tab2}, row 9.

In the first step, the source, $S$, is imaged with the collimating
mirror (lens) onto an intermediate reference plane at distance
$l_{\ind{1}}$ from the mirror. The image is calculated using the
focusing system ray-transfer matrix
$\hat{\focus}(l_{\ind{1}},f_{\ind{1}},f_{\ind{1}})$ with the
assumption that the source is placed at the focal distance,
$f_{\ind{1}}$, from the collimating mirror.  In the second step,
transformations by the crystal optic (dispersing element of the
spectrograph) are described by the ray-transfer matrix
$\hat{\crystal}(\dcomm{n} ,\fcomm{n})$. We assume at this point that
the distances between the crystals are negligible. In the third step,
the focusing mirror (lens) with a focal length $f_{\ind{2}}$ placed at
distance $l_{\ind{2}}$ from the crystal system produces the source
image in the focal plane, as described by the ray-transfer matrix
$\hat{\focus}(f_{\ind{2}},f_{\ind{2}},l_{\ind{2}})$. The final source
image is described by a spectrograph matrix that is a product of the
tree matrices $\hat{\focus}(f_{\ind{2}},f_{\ind{2}},l_{\ind{2}})
\hat{\crystal}(\dcomm{n} ,\fcomm{n})
\hat{\focus}(l_{\ind{1}},f_{\ind{1}},f_{\ind{1}}) $ from
Table~\ref{tab2}.  The spectrograph ray-transfer matrix
$\hat{\spectro}(f_{\ind{2}},l_{\ind{2}},\dcomm{n}\!,\fcomm{n}\!,l_{\ind{1}},f_{\ind{1}})$
in given in row 9 of Table~\ref{tab2}.

Remarkably, element $B$ of the spectrograph matrix is zero. This means
that for a monochromatic light the spectrograph is working as a
focusing system, concentrating all photons from a point source into a
point image, independent of the initial angular size of the source.
Using matrix element $A$, we calculate that the spectrograph projects a
monochromatic source with a linear size $\Delta x$ into an image of
linear size
\begin{equation}
  \Delta x^{\prime} =  \dcomm{n}  \frac{f_{\ind{2}}}{f_{\ind{1}}} \Delta x.
\label{ssr010}
\end{equation}
If the source is not monochromatic, the source image produced by the
photons with energy $E+\delta E$ is shifted transversely due to linear
dispersion by
\begin{equation}
\delta x^{\prime} = f_{\ind{2}} \fcomm{n} \delta E 
\label{ssr020}
\end{equation}
from the source image by photons with energy $E$. The spectrograph
spectral resolution, $\Delta E$, can be determined from the condition
that the monochromatic source image size $\Delta x^{\prime}$ 
[Eq.~\eqref{ssr010}], is equal to the source image shift $\delta
x^{\prime}$ [Eq.~\eqref{ssr020}]:
\begin{equation}
\Delta E =  \frac{\Delta x}{f_{\ind{1}}} \frac{| \dcomm{n} |}{\fcomm{n} }. 
\label{ssr030}
\end{equation}
A large cumulative dispersion  rate $\fcomm{n}$, a small cumulative
asymmetry factor $|\dcomm{n}|$, a large focal distance $f_{\ind{1}}$ of the
collimating mirror, and a small source size $\Delta x$ are
advantageous for better spectral resolution. 

Comparing Eq.~\eqref{ssr030} with Eqs.~\eqref{fm140} and
\eqref{fm230}, we note that the spectral resolution of the focusing
monochromators and of the spectrograph are described by the same
expressions, with the only difference being that the source-lens
distance is $f_{\ind{1}}$ in the case of the spectrograph, and
$l_{\ind{1}}$ in the case of the monochromators. We therefore reach an
interesting conclusion: the spectral resolution of the focusing
monochromators and spectrographs can be equivalent. However, their
angular acceptance and spectral efficiency, may be substantially
different.

The ray-transfer matrix theory does not take into account spectral and
angular widths of the Bragg reflections involved.  They are, however,
often limited typically to relatively small eV--meV spectral and to
mrad--$\mu$rad angular widths. The collimating optic of the
spectrograph produces a beam with an angular divergence $\Delta
x/f_{\ind{1}}$ from a source with a linear size of $\Delta x$
(independent of the angular size of the source). If $\Delta
x/f_{\ind{1}}$ is chosen to be smaller than the angular acceptance of
the crystal optic, the spectrograph may accept photons from a source
with a large angular size.  The focusing monochromators, which use
only one lens (mirror) in their optic, do not have such adaptability
to sources with large angular size. Focusing monochromators can work
efficiently only with sources of small angular size, smaller than the
angular acceptance of the crystal optic. Therefore, spectrographs are
preferable spectral imaging systems to work with sources of large
angular size. This is exactly the requirement for the analyzer systems
of the IXS instruments. In the following sections we will therefore
consider only spectrographs in application to IXS.

The spectrograph ray-transfer matrix $\hat{\spectro}$ presented in
Table~\ref{tab2} was derived neglecting propagation through free space
between the crystals. It turns out that only matrix elements $C$ and
$F$ have to be changed if nonzero distances between the crystals of
the spectrograph are taken into account:
\begin{equation}
\tilde{C}= C+\Delta C, \hspace{1cm} \Delta C = - \frac{\bcomm{n}}{f_{\ind{1}}f_{\ind{2}}},
\end{equation}
\begin{equation}
\tilde{F}= F+\Delta F, \hspace{1cm} \Delta F = - \frac{\gcomm{n}}{f_{\ind{2}}}.
\end{equation}
However, this leaves intact the results of the analysis presented
above, because these elements were not used to derive
Eqs.~\eqref{ssr010}--\eqref{ssr030}.

\section{Broadband Spectrographs}
\label{broadband-spectrographs}

A perfect x-ray imaging spectrograph for IXS applications should have
a high spectral resolution, $\Delta E$ ($\Delta E/E\ll 10^{-6}$); a
large spectral window of imaging, $\Delta E_{\ind{\cup}} \gg \Delta E$;
and a large angular acceptance, $\Delta \theta_{\ind{\cup}} \simeq 1-10$~mrad.

\CT-type spectrographs are large-acceptance-angle devices in contrast
to focusing monochromators, as discussed in detail in
Section~\ref{spectrograph-section}.  Therefore, in this section we
will consider \CT-type spectrographs as spectral imaging systems for
IXS spectroscopy.

To achieve required spectral resolution $\Delta E$, the ``diffraction
grating'' parameters, $\dcomm{n}$ and $\fcomm{n}$; the focal length,
$f_{\ind{1}}$, of the collimating optic; and the source size, $\Delta x$,
have to be appropriately selected using Eq.~\eqref{ssr030}. We will
discuss this in more detail later in this section.

The key problem is how to achieve large spectral window of imaging
$\Delta E_{\ind{\cup}} \gg \Delta E$ (i.e., how to achieve broadband
spectrographs).  In the ray-transfer matrix theory presented above,
infinite reflection bandwidths of the optical elements have been
assumed. In reality, Bragg reflection bandwidths are narrow. They are
determined in the dynamical theory of x-ray diffraction in perfect
crystals (see, e.g., \cite{Authier,Shvydko-SB}). Therefore, we have to
join ray-transfer matrix approach with the dynamical theory to
tackle the problem of the spectrograph bandwidth.

In the following sections, we will consider two types of multi-crystal
dispersing elements that may be used as ``diffraction gratings'' of
the broadband hard x-ray spectrographs with very high spectral
resolution.

\subsection{0.1-meV resolution broadband spectrographs with CDW
  dispersing elements}
 \label{cdw-uhrix}

\begin{figure*}[t!]
\setlength{\unitlength}{\textwidth}
\begin{picture}(1,0.46)(0,0)
\put(0.0,0.00){\includegraphics[width=1.0\textwidth]{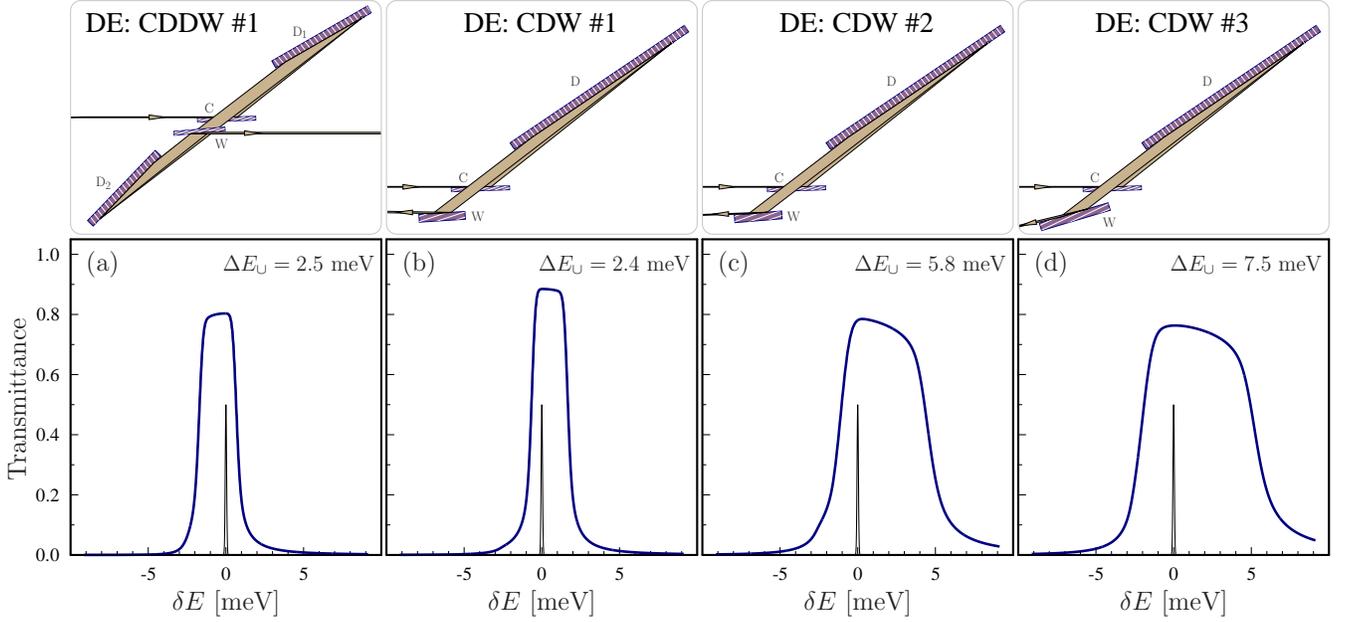}}
\end{picture}
\caption{(Color online) Schematics of the four-crystal CDDW and three-crystal CDW optic as
  dispersing elements (``diffraction gratings'') of the \CT -type hard
  x-ray imaging spectrographs (top row) and their spectral
  transmittance functions (solid, dark blue lines in the bottom row)
  calculated by the dynamical theory of Bragg diffraction using
  crystal parameters from Table~\ref{tab1}.  Angular spread of
  incident x-rays is $20~\mu$rad in (a)--(b), and $50~\mu$rad in
  (c)--(d). Black spectral lines  with a 0.1-meV
  width indicate the target spectral resolution of the spectrographs.}
\label{fig0012}
\end{figure*}

 \CT -type hard x-ray spectrographs using the CDW optic
 \cite{Shvydko-SB,SLK06,SKR06,ShSS11} as the dispersing element has
 been introduced in \cite{Shv11,Shvydko12,SSM13}.  Three-crystal
 CDW-optic schematics are shown in
 Figs.~\ref{fig0012}(b)-\ref{fig0012}(c), while Fig.~\ref{fig0012}(a)
 shows its four-crystal modification CDDW comprising two D-crystal
 elements.

The CDW optic in general and CDDW optic in particular may feature the
cumulative dispersion rates, $\fcomm{}$, greatly enhanced by successive
asymmetric Bragg reflections. The enhancement is described by the
equation from row 4 of Table~\ref{tab2}:
\begin{equation}
\fcomm{n}=b_{\ind{n}}\fcomm{n-1} + \sgn_{\ind{n}}\dirate_{\ind{n}}.
\label{ssr080}
\end{equation}
It tells that the dispersion rate $\fcomm{n-1}$ of the optic composed
of the first $n-1$ crystals can be drastically enhanced, provided
successive crystal's asymmetry factor $|b_{\ind{n}}|\gg 1$. In the
example discussed in \cite{Shv11,Shvydko12,SSM13}, the CDDW optic was
considered, for which the cumulative dispersion rate was enhanced
almost by two orders of magnitude compared to that of a single Bragg
reflection. As a consequence, the ability to achieve very high
spectral resolution $\Delta E < 0.1$~meV was demonstrated. However,
the spectral window in which that particular CDDW optic permitted the
imaging of x-ray spectra was only $\Delta E_{\ind{\cup}} \simeq
0.45$~meV.

Here we introduce x-ray spectrographs with the dispersing elements
using the CDW optic, which feature a more than an order-of-magnitude
increase (compared to the \cite{SSM13} case) in the spectral window of
imaging, and simultaneously a very high spectral resolution $\Delta E
\simeq 0.1$~meV.

A spectrograph with a spectral resolution $\Delta E=0.1$~meV requires
a dispersing element (DE in Fig.~\ref{fig001}), featuring the ratio
$|\dcomm{}|/|\fcomm{}|=0.02$~meV/$\mu$rad [see
Eq.~\eqref{ssr030}]. Here we assume that the source size on the sample
$\Delta x=5~\mu$m, and focal distance $f_{\ind{1}}=1$~m. Small
$|\dcomm{}|$ and large $|\fcomm{}|$ are favorable. However, a
$|\dcomm{}|$ value that is too small may result in an enlargement by
$1/|\dcomm{}|$ of the transverse size of the beam after the dispersing
element, which is a too big, and therefore may require focusing optic
with unrealistically large geometrical aperture. In addition, a
$|\dcomm{}|$ value that is too small may result in a monochromatic
image size $\Delta x^{\prime}$ that is too small [see
Eq.~\eqref{ssr020}], which may be beyond the detector's spatial
resolution.  Because of this, we will keep $|\dcomm{n}|\simeq 0.5$;
therefore, $|\fcomm{}|\simeq 25~\mu$rad/meV in the examples considered
below. With $f_{\ind{2}}/f_{\ind{1}}\simeq 1-2$, the monochromatic
image size is expected to be $\Delta x^{\prime} \simeq 2.5-5 ~\mu$m,
which can be resolved by  modern position-sensitive x-ray detectors
\cite{SBD12}.  It is also important to ensure that the angular acceptance
of spectrograph's dispersing element is much larger than the angular size
of the source $\simeq \Delta x/f_{\ind{1}}\simeq 5~\mu$rad.

\begin{table}[t!]
\centering
\begin{tabular}{|l|lllllll|}
  \hline   \hline 
crystal   & $\vc{H}_{\indrm{\elmt}}$ &$\eta_{\indrm{\elmt}} $ &$\theta_{\indrm{\elmt}} $  & $\dei{\elmt} $ &  $\dai{\elmt}$  & $b_{\indrm{\elmt}}$ & $s_{\indrm{\elmt}}\dirate_{\indrm{\elmt}} $ \\[-5pt]    
element (\elmt)  & &  &  &   &  &   &    \\[0pt]    
[material]          & $(hkl)$ & deg & deg  & meV  &  $\mu$rad   & & $\frac{\mu {\mathrm {rad}}}{\mathrm {meV}}$ \\[5pt]    
\hline  \hline  
\multicolumn{8}{|l|}{CDDW \#1}\\  
\cline{1-1}
  C~~[C*]& (1~1~1) &  -17.3  &  19.26  &  574 & 22    & -0.057  & -0.03 \\[-0.0pt]
  D$_{\ind{1}}$~[Si] & (8~0~0) &  81.9  &  89.5  &  27 &  341    & -1.13  & 1.63 \\[-0.0pt]
  D$_{\ind{2}}$~[Si] & (8~0~0) &  81.9  &  89.5   & 27 &  341   & -1.13  & -1.63 \\[-0.0pt]
  W~~[C*] & (1~1~1) &  14.6  &  19.25   & 574 &  22   & -6.88  & 0.22 \\[-0.0pt]
  \hline  
\multicolumn{4}{|l}{}  & $\Delta E_{\ind{\cup}}$  & $\Delta \theta_{\ind{\cup}}^{\prime}$  & $\dcomm{ }$ & $\fcomm{ }$ \\
\multicolumn{4}{|l}{}  & meV  & $\mu$rad  &  & $\frac{\mu {\mathrm {rad}}}{\mathrm {meV}}$ \\[5pt]    
  \hline  
\multicolumn{4}{|l}{Cumulative values}  & 2.5  & 62  & 0.5 & 24.6 \\
\hline  \hline  
\multicolumn{8}{|l|}{CDW \#1}\\  
\cline{1-1}
  C~[C*]& (1~1~1) &  -17.3  &  19.26  &  574 & 22    & -0.057  & -0.03 \\[-0.0pt]
  D~[Si] & (8~0~0) &  86.0  &  89.5   & 27 &  341   & -1.29  & 3.58 \\[-0.0pt]
  W~[C*] & (1~1~1) &  14.55  &  19.25   & 574 &  22   & -6.79  & -0.22 \\[-0.0pt]
  \hline  
\multicolumn{4}{|l}{Cumulative values}  & 2.4  & -60  & -0.5 & -24.9 \\
  \hline  
  \hline  
\multicolumn{8}{|l|}{CDW \#2}\\  
\cline{1-1}
  C~[C*]& (1~1~1) &  -17.3  &  19.26  &  574 & 22    & -0.057  & -0.03 \\[-0.0pt]
  D~[Si] & (8~0~0) &  86.0  &  89.5   & 27 &  341   & -1.29  & 3.58 \\[-0.0pt]
  W~[Ge] & (2~2~0) &  15.0  &  19.84   & 1354 &  53   & -6.77  & -0.22 \\[-0.0pt]
  \hline  
\multicolumn{4}{|l}{Cumulative values}  & 5.8  & -144  & -0.5 & 24.8 \\
  \hline  
  \hline  
\multicolumn{8}{|l|}{CDW \#3}\\  
\cline{1-1}
  C~[C*]& (1~1~1) &  -17.3  &  19.26  &  574 & 22    & -0.057  & -0.03 \\[-0.0pt]
  D~[Si] & (8~0~0) &  86.0  &  89.5   & 27 &  341   & -1.29  & 3.58 \\[-0.0pt]
  W~[Ge] & (1~1~1) &  9.0  &  12.0   & 3013 &  70   & -6.86  & -0.13 \\[-0.0pt]
  \hline  
\multicolumn{4}{|l}{Cumulative values}  & 7.5  & -187  & -0.5 & 25. \\
  \hline  
  \hline  
\end{tabular}
\caption{Examples of the four-crystal CDDW and three-crystal CDW optic as dispersing elements 
  (``diffraction gratings'') of the \CT -type hard x-ray spectrographs. 
  For each optic the table presents crystal elements (\elmt=C,D,W) and their Bragg reflection parameters:
  $(hkl)$, Miller indices of the Bragg diffraction vector $\vc{H}_{\indrm{\elmt}}$;  
  $\eta_{\indrm{\elmt}}$, asymmetry angle; $\theta_{\indrm{\elmt}}$, glancing angle of incidence; $\deis{\elmt}$, $\dais{\elmt}$
  are Bragg's reflection intrinsic spectral width and angular acceptance in symmetric scattering geometry, respectively; 
  $b_{\indrm{\elmt}}$, asymmetry factor; and  $s_{\indrm{\elmt}} \dirate_{\indrm{\elmt}} $, angular dispersion rate with deflection sign.  For each optic the table also shows  the 
  spectral window of imaging $\Delta E_{\ind{\cup}}$ as derived from the dynamical  theory calculations - see Fig.~\ref{fig0012}, the angular spread of the dispersion fan 
  $\Delta \theta_{\ind{\cup}}^{\prime}=\fcomm{ } \Delta E_{\ind{\cup}}$, 
  and cumulative values of asymmetry parameter $\dcomm{ }$  and dispersion rate $\fcomm{ }$.
  X-ray photon energy is $E=9.13185$~keV in all cases.
}
\label{tab1}
\end{table}

Based on the DuMond diagram analysis, the spectral bandwidth of the
CDDW optic can be approximated by the following expression \cite{ShSS11}:
\begin{equation}
\Delta E_{\ind{\cup}} \simeq E \frac{\Delta\theta^{({\mathrm s})}_{\indrm{C}} \sqrt{|b_{\indrm{C}}|}  + \Delta\theta^{({\mathrm s})}_{\indrm{W}}/\sqrt{|b_{\indrm{W}}|}  }{4 \tan\eta_{\indrm{D}}}. 
\label{ssr052}
\end{equation}
Here $\Delta\theta^{({\mathrm s})}_{\indrm{\elmt}}$ values represent
angular widths of Bragg reflections from crystal elements (\elmt\ =
C,W) in the symmetric scattering geometry.

For the CDDW optic, $s_{\indrm{C}}=+1$; $s_{\indrm{D1}}=+1$;
$s_{\indrm{D2}}=-1$; and $s_{\indrm{W}}=-1$. Therefore, using
Eq.~\eqref{ssr080}, 
\begin{equation}
\fcomm{} =
b_{\indrm{W}}b_{\indrm{D2}}b_{\indrm{D1}}\dirate_{\indrm{C}}+b_{\indrm{W}}b_{\indrm{D2}}\dirate_{\indrm{D1}}-b_{\indrm{W}}\dirate_{\indrm{D2}}-\dirate_{\indrm{W}}. 
\label{ssr057}
\end{equation}
Assuming
typical designs with
$\theta_{\indrm{D1}}=\theta_{\indrm{D2}}=\theta_{\indrm{D}} \simeq
90^{\circ}$, and $b_{\indrm{D1}}=b_{\indrm{D2}}=b_{\indrm{D}} \simeq
-1$, the largest dispersing rates are achieved by D-crystals,
$\dirate_{\indrm{D1}} = \dirate_{\indrm{D2}} =\dirate_{\indrm{D}} =
2\tan\eta_{\indrm{D}} /E $ [see Eq.~\eqref{ad010}], while the
dispersion rates $\dirate_{\indrm{C}}$ and $\dirate_{\indrm{W}}$ of
the C- and W-crystal elements can be neglected in Eq.~\eqref{ssr057}.
As a result, the cumulative dispersion rate can be then approximated by
\begin{equation}
\fcomm{} \simeq  -2 b_{\indrm{W}}\dirate_{\indrm{D}} \simeq -4 b_{\indrm{W}}\tan\eta_{\indrm{D}} /E ,
\label{ssr050}
\end{equation}
and the critical for spectrograph's spectral resolution $\Delta E$
ratio $\dcomm{}/\fcomm{}$ [see Eq.~\eqref{ssr030}] by
\begin{equation}
\frac{ \dcomm{} }{\fcomm{}} \simeq -E \frac{b_{\indrm{C}}b_{\indrm{D}}^2}{4\tan\eta_{\indrm{D}}}.
\label{ssr056}
\end{equation}
Equation \eqref{ssr052} shows that to achieve a broadband spectrograph
it is important to use the W-crystal with a large intrinsic angular
width $\Delta\theta^{({\mathrm s})}_{\indrm{W}}$ and a small asymmetry
factor $|b_{\indrm{W}}|$; however, asymmetry factor should not be too
small, in order to keep $|\dcomm{}|\simeq 0.5$, as discussed
above. Favorably, the variation of $|b_{\indrm{W}}|$ does not change
the spectral resolution $\Delta E$, according to Eqs.~\eqref{ssr056}
and \eqref{ssr030}.  Using the C-crystal with a large intrinsic
angular width $\Delta\theta^{({\mathrm s})}_{\indrm{C}}$, and as small
as possible asymmetry factor $|b_{\indrm{C}}|$ is also advantageous
for achieving the large bandwidth.  However, these values are
optimized first of all with a purpose of achieving a large angular
acceptance of the CDDW optic $\Delta \theta_{\ind{\cup}} \simeq
\Delta\theta^{({\mathrm s})}_{\indrm{C}}/ \sqrt{|b_{\indrm{C}}|}$
\cite{ShSS11}.

Equations~\eqref{ssr052}--\eqref{ssr056} are also valid for the
three-crystal CDW optic, if factors of $4$ are replaces everywhere by
factors of $2$, and $b_{\indrm{D}}^2$ is replaced by
$b_{\indrm{D}}$. Therefore, similar conclusions regarding the
bandwidth are true for the CDW optic, containing one D-crystal
element.

Examples of multi-crystal CDDW and CDW ``diffraction gratings'',
ensuring $\Delta E_{\ind{\cup}} \simeq 2.5-7.5$~meV (i.e., spectral
windows of imaging 25 to 75 times broader than the target spectral
resolution $\Delta E=0.1$~meV) are given in Table~\ref{tab1}. The
spectral transmittance functions calculated using the dynamical
diffraction theory are shown in Fig.~\ref{fig0012}. The largest
increase in the width of the spectral window of imaging is achieved in
those cases in which Ge crystals are used for W-crystal elements, the
crystals that provide the largest $\Delta\theta^{({\mathrm
    s})}_{\indrm{W}}$ values.  Low-indexed asymmetric Bragg
reflections from thin diamond crystals, C*, are proposed to use for
the C-crystal elements, to ensure low absorption of the beam
propagating to the W-crystal upon Bragg back-reflection from the
D-crystal, similar to how diamond crystals were used in a hybrid
diamond-silicon CDDW x-ray monochromator \cite{SSS13}.

The above examples are not necessarily best and final.  Further
improvements in the spectral resolution and the spectral window of
imaging are still possible through changing crystal parameters and
crystal material.  The best strategy of increasing the spectral window
of imaging is to choose a crystal material with the largest possible
angular acceptance, $\Delta\theta^{({\mathrm s})}_{\indrm{W}}$, of the
W-crystal [Eq~\eqref{ssr052}]. Here we suggest Ge, but a different
material could also work (e.g., PbWO$_4$).  The spectral window of
imaging can be further increased by decreasing the asymmetry factor,
$|b_{\indrm{W}}|$, of the W-crystal while simultaneously keeping
$\dcomm{}/\fcomm{}$, and therefore $\Delta E$, at the same low level.
This should be possible as long as the transverse size of the beam
which has been increased by $1/|\dcomm{}|$ after the CDW optic can be
accepted by the focusing optic, and the monochromatic image size
$\Delta x^{\prime}$ decreased by $|\dcomm{}|$ [see Eq.~\eqref{ssr010}]
can be still resolved by the detector.

In addition, with the Bragg angle, $\theta_{\indrm{D}}$, of the
D-crystal chosen very close to $90^{\circ}$ the CDW optic becomes
exact back-scattering. In this case, a Littrow-type spectrograph in an
autocollimating configuration, using common crystal for C- and W-
elements, could be used. This is similar to the \CT -type spectrograph
but with a common collimator and focusing mirror.

\subsection{Spectrographs with a  multi-crystal  (+--+--...) dispersing element}

In the asymmetric scattering geometry with angle $\eta \not = 0$ [see
Fig.~\ref{fig001}(b)] the relative spectral width $\Delta E/E$ and
angular width $\Delta \theta$ of the Bragg reflection region become
\begin{equation}
\frac{\Delta E}{E}\,=\,\frac{\ehs}{\sqrt{|b|}}, \hspace{0.5cm}\Delta\theta\,=\, \frac{\Delta\theta^{({\mathrm s})}}{\sqrt{|b|}},
\label{ssr090}
\end{equation}
compared to the appropriate values $\ehs$ and $\Delta\theta^{({\mathrm
    s})}\simeq \ehs \tan\theta $ valid in the symmetric scattering
geometry with $\eta=0$ and $b=-1$ (see, e.g.,
\cite{Authier,Shvydko-SB}).  The spectral and angular Bragg reflection
widths increase by a factor $1/\sqrt{|b|}$ compared to symmetric case
values, provided Bragg reflections with asymmetry parameters $|b|<1$
($\eta < 0$) are used.  It is therefore clear that to realize
spectrographs with broadest possible spectral window of imaging it is
advantageous to use asymmetric Bragg reflections with asymmetry
parameters in the range of $|b|<1$. In the current section, we will
study a few particular cases.

We start, however, with drawbacks of using Bragg reflections with
asymmetry factors in the range of $|b|<1$. First, they have a
smaller angular dispersion rates [see Eq.~\eqref{ad010}] than those
with $|b|\gg 1$ (compare with cases discussed in the previous
section).  Second, they enlarge the transverse beam size of x-rays
upon reflection by a factor $1/|b|$, as a consequence of the phase
space conservation (see $A$ matrix elements of the Bragg reflection
ray-transfer matrices in rows 3--5 of Table~\ref{tab2}).  Transverse
beam sizes that are too large may unfortunately require
unrealistically large geometric aperture of the $f_{\ind{2}}$-focusing
mirrors (lenses) of the spectrographs.  Third, Bragg reflections with
$|b|\ll 1$ also reduce the monochromatic image size $\Delta
x^{\prime}$ [see Eq.~\eqref{ssr010}] and thus may pose stringent
requirement on the detector's spatial resolution, which should be
better than $\Delta x^{\prime}$.

\begin{figure}[t!]
\setlength{\unitlength}{\textwidth}
\begin{picture}(1,0.33)(0,0)
\put(0.0,0.00){\includegraphics[width=0.5\textwidth]{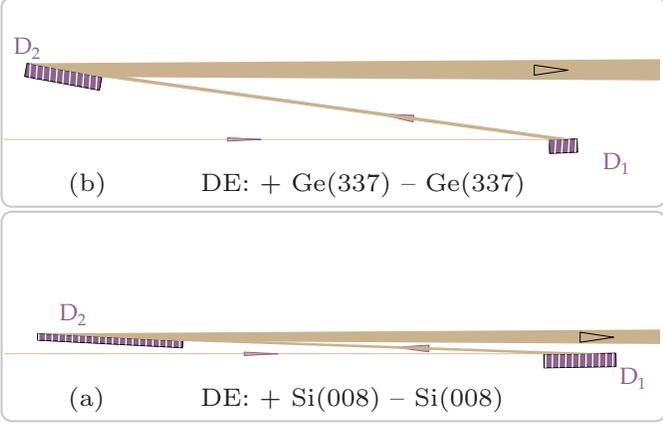}}
\end{picture}
\caption{(Color online) Schematics of two-crystal optic in the (+--) arrangement
  designed as dispersing elements DE (``diffraction gratings'') of
  the \CT -type hard x-ray imaging spectrographs
  (Fig~\ref{fig001}). Identical Bragg reflections with asymmetry
  parameter $|b|<1$ are used, with crystal parameters presented in
  Tables~\ref{tab3}(a) and \ref{tab3}(b), and with spectral transmittance
  functions presented in Figs.~\ref{fig0017}(a) and \ref{fig0017}(b),
  respectively. }
\label{fig008}
\end{figure}

As  example, we consider a multi-crystal dispersing element of the
\CT -type spectrographs composed of $n$ identical crystals in the
(+--+--...)  scattering geometry ($s_{\ind{1}}=+1$, $s_{\ind{2}}=-1$,
$s_{\ind{3}}=+1$, $s_{\ind{4}}=-1$,...). All crystals are assumed to
have the same angular dispersion rate
$\dirate_{\ind{1}}=\dirate_{\ind{2}}= ... = \dirate_{\ind{n}}=\dirate$
[see Eq.~\eqref{ad010}], and the same asymmetry factors
$b_{\ind{1}}=b_{\ind{2}}= ... = b_{\ind{n}}=b$ [see
Eq.~\eqref{ad090}]. Using the equations from row 4 of
Table~\ref{tab2}, we obtain for the cumulative dispersion rate
$\fcomm{n}$, and the asymmetry factor $\dcomm{n}$ of the multi-crystal
(+--+--...)  dispersing element:
\begin{gather}
\label{ssr040}
\fcomm{n}=\dirate s_{\ind{n}}
(1-b+b^2+...+s_{\ind{n}}b^{n-1}),\\
\dcomm{n}=b^n.
\label{ssr045}
\end{gather}
Increasing the number of crystals, $n$, with asymmetry parameters
$|b|<1$, results in a rapid decrease of $\dcomm{n}$; however, this
does not increase $\fcomm{n}$ as much. Therefore, in the following
examples we restrict ourselves to considering solely two-crystal
(+--)-type dispersing elements, as shown schematically in
Figure~\ref{fig008}.

For a spectrograph with two-crystal (+--)-type dispersing elements,
the expressions for the spectral resolution $\Delta E$ [see
Eqs.~\eqref{ssr030} and \eqref{ssr040}-\eqref{ssr045}] and for the
monochromatic image size $\Delta x^{\prime}$ [see Eqs.~\eqref{ssr010}
and \eqref{ssr045}] on the detector become
\begin{equation}
  \Delta E = \frac{\Delta x}{f_{\ind{1}}} \frac{b^2}{\dirate (b-1)} = \frac{\Delta x}{f_{\ind{1}}} E \frac{b^2}{(1-b^2)\tan\theta}
\label{ssr060}
\end{equation}
\begin{equation}
  \Delta x^{\prime} =  b^2  \frac{f_{\ind{2}}}{f_{\ind{1}}} \Delta x.
\label{ssr070}
\end{equation}
Bragg reflections with $\theta$ close to $90^{\circ}$, small $|b|\ll
1$, small source size $\Delta x$, and large focal distance,
$f_{\ind{1}}$, are the factors that improve the energy resolution
$\Delta E$ of the spectrograph. A large focal distance, $f_{\ind{2}}$,
of the spectrograph's focusing mirror helps to mitigate the
requirement for the spatial resolution of the position sensitive
detector.

The spectral window of imaging $\Delta E_{\ind{\cup}}$ and the angular
acceptance $\Delta \theta_{\ind{\cup}}$ of the spectrograph for each
monochromatic spectral component is given by Eq.~\eqref{ssr090} in the
first approximation. Using Bragg reflections with a large relative
spectral widths, $\ehs$, and a small $|b|\ll 1$ is advantageous for
achieving a broad spectral window of imaging.

In the following, we consider two examples of the spectrographs in the
\CT\ configuration with two-crystal (+--)-type dispersing elements.
The first one, which is appropriate for UHRIXS applications, is
studied in Section~\ref{two-cristal-uhrix}. The second example,
relevant to high-resolution Cu K-edge RIXS applications, is discussed
in Section~\ref{two-cristal-rixs}.

\subsubsection{IXS spectrograph:  0.1-meV resolution and 47-meV spectral window}
\label{two-cristal-uhrix}

\begin{figure}[t!]
\setlength{\unitlength}{\textwidth}
\begin{picture}(1,0.28)(0,0)
\put(0.0,0.00){\includegraphics[width=0.5\textwidth]{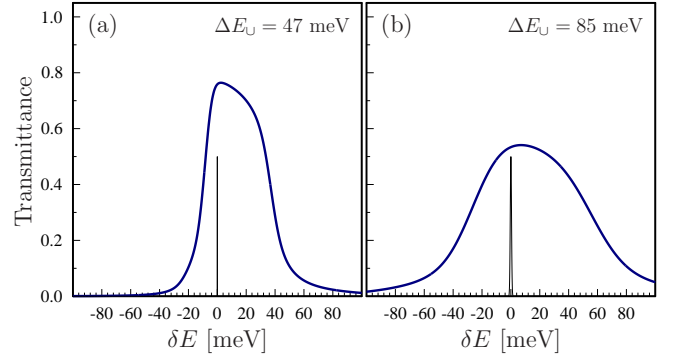}}
\end{picture}
\caption{(Color online) Spectral transmittance functions [solid dark blue lines in (a)
  and (b)] of the (+--)-type two-crystal dispersing elements DE,
  schematically shown in Fig.~\ref{fig008}(a) and (b), respectively.
  Transmittance is calculated using the dynamical theory of Bragg
  diffraction with crystal parameters from Tables~\ref{tab3}(a) and
  \ref{tab3}(b).  Angular spread of incident x-rays is $50~\mu$rad in both
  cases. Black spectral lines with a 0.1-meV width in (a) and with a
  1-meV width in (b) represent the spectral resolution of the
  spectrographs in the particular $(\theta,\eta)$ configurations
  highlighted by magenta dots in Figs.~\ref{fig009} and
  \ref{fig009899}, respectively.}
\label{fig0017}
\end{figure}

\begin{figure}[t!]
\setlength{\unitlength}{\textwidth}
\begin{picture}(1,1.0)(0,0)
\put(0.0,-0.02){\includegraphics[width=0.5\textwidth]{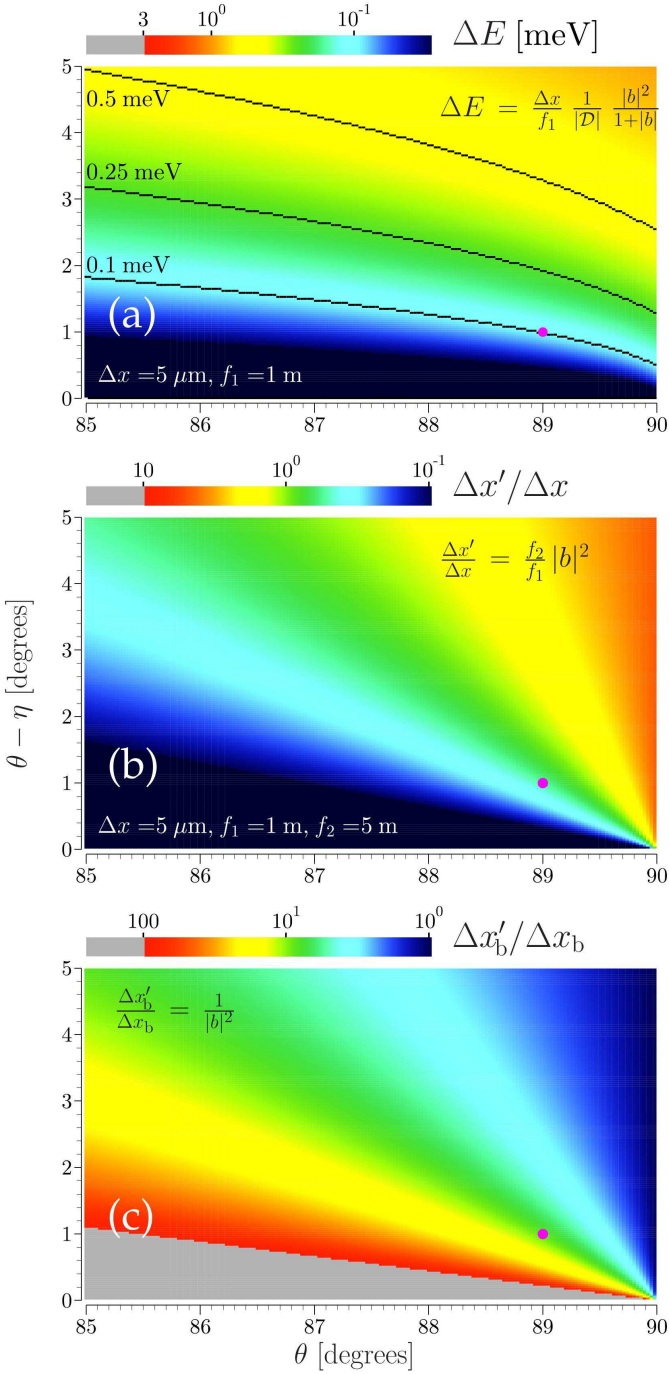}}
\end{picture}
\caption{Properties of the spectrographs with the two-crystal ``diffraction
  grating'' (Fig.~\ref{fig008}): (a) spectral resolution $\Delta E$;
  (b) image to source size ratio $\Delta x^{\prime}/\Delta x$; and (c)
  the lateral beam size  enlargement $\Delta \sz^{\prime}/\Delta \sz$ by
  crystal optics, shown as a function of Bragg's angle $\theta$ and
  the asymmetry angle $\eta$.  A particular case is presented for
  $\Delta x=5~\mu$m, $f_{\ind{1}}=1$~m, and $f_{\ind{2}}=5$~m. Magenta
  dots highlight the case with the spectrograph resolution $\Delta
  E=0.1$~meV, $\Delta x^{\prime}/\Delta x \simeq 0.55$, and $\Delta
  \sz^{\prime}/\Delta \sz \simeq 9$, attainable with crystal parameters
  presented in Table~\ref{tab3}(a). The (008) Bragg reflections from
  Si crystals enable a ``diffraction grating'' with a $\Delta \theta_{\ind{\cup}} =
  284~\mu$rad angular acceptance and $\Delta E_{\ind{\cup}}=47$~meV
  imaging window for $9.13294$~keV x-rays.  }
\label{fig009}
\end{figure}

\begin{table}[t!]
\centering
\begin{tabular}{|l|lllllll|}
  \hline   \hline 
crystal   & $\vc{H}_{\indrm{\elmt}}$ &$\eta_{\indrm{\elmt}} $ &$\theta_{\indrm{\elmt}} $  & $\dei{\elmt} $ &  $\dai{\elmt}$  & $b_{\indrm{\elmt}}$ & $s_{\indrm{\elmt}}\dirate_{\indrm{\elmt}} $ \\[-5pt]    
element (\elmt)  & &  &  &   &  &   &    \\[0pt]    
[material]          & $(hkl)$ & deg & deg  & meV  &  $\mu$rad   & & $\frac{\mu {\mathrm {rad}}}{\mathrm {meV}}$ \\[5pt]    
\hline  \hline  
\multicolumn{8}{|l|}{(a) $E=9.13294$~keV}\\  
\cline{1-1}
  D$_{\ind{1}}$~[Si] & (8~0~0) &  88.  &  89.  &  27 &  169   & -0.34  & -4.2 \\[-0.0pt]
  D$_{\ind{2}}$~[Si] & (8~0~0) &  86.  &  89.  & 27 &  169   & -0.34  &  4.2 \\[-0.0pt]
  \hline  
\multicolumn{4}{|l}{}  & $\Delta E_{\ind{\cup}}$  & $\Delta \theta_{\ind{\cup}}^{\prime}$  & $\dcomm{ }$ & $\fcomm{ }$ \\
\multicolumn{4}{|l}{}  & meV  & $\mu$rad  &  & $\frac{\mu {\mathrm {rad}}}{\mathrm {meV}}$ \\[5pt]    
  \hline  
\multicolumn{4}{|l}{Cumulative values}  & 47  & 266  & 0.11 & 5.63 \\
\hline  \hline  
\multicolumn{8}{|l|}{(b) $E=8.99$~keV}\\  
\cline{1-1}
  D$_{\ind{1}}$~[Ge] & (3~3~7) &  83.45  &  86.05  &  41.8 &  67    & -0.25  & -1.2 \\[-0.0pt]
  D$_{\ind{2}}$~[Ge] & (3~7~7) &  83.45  &  86.05   & 41.8 &  67   & -0.25  & +1.2 \\[-0.0pt]
  \hline  
\multicolumn{4}{|l}{Cumulative values}  & 85  & 134  & -0.06 & 1.5 \\
  \hline  
  \hline  
\end{tabular}
\caption{Examples of the two-crystal (+--)-type optic designed 
  as dispersing elements DE (``diffraction gratings'') of the \CT -type 
  hard x-ray imaging spectrographs. All notations are as in Table~\ref{tab1}.}
\label{tab3}
\end{table}

Here, as in Section~\ref{cdw-uhrix}, we study possible solutions to
broadband spectrographs for IXS applications that require an
ultra-high spectral resolution of $\Delta E \simeq 0.1$~meV, and a
momentum transfer resolution $\Delta Q \simeq 0.01$~nm$^{-1}$
discussed in Section~\ref{intro}.

We use Eqs.~\eqref{ssr060}--\eqref{ssr070} and \eqref{ad090} to plot
two-dimensional (2D) graphs with spectrograph characteristics as a
function of Bragg's angle $\theta$ and the asymmetry angle $\eta$:
spectral resolution $\Delta E$ in Fig.~\ref{fig009}(a), image to
source size ratio $\Delta x^{\prime}/\Delta x$ in
Fig.~\ref{fig009}(b), and the lateral beam size enlargement
$\Delta \sz^{\prime}/\Delta \sz$ by crystal optics in
Fig.~\ref{fig009}(c).  A particular case is considered with $\Delta
x=5~\mu$m, $f_{\ind{1}}=1$~m, and $f_{\ind{2}}=5$~m.  Configurations
with equal energy resolution are highlighted by black lines for some
selected $\Delta E$ values. Magenta dots highlight a specific case
with the spectral resolution $\Delta E=0.1$~meV, $\Delta
x^{\prime}/\Delta x \simeq 0.55$, and $\Delta \sz^{\prime}/\Delta \sz
\simeq 9$, achieved by selecting $\theta=89^{\circ}$ and
$\theta-\eta=1^{\circ}$ ($b=-0.33$). Specifically, the (008) Bragg
reflection of x-rays with average photon energy $E=9.13294$~keV from
Si crystals [see Table~\ref{tab3}(a), Figs.~\ref{fig0017}(a) and
\ref{fig008}(a)] enable a ``diffraction grating'' with a spectral
window of imaging $\Delta E_{\ind{\cup}}=47$~meV and angular
acceptance $\Delta \theta_{\ind{\cup}}= 266~\mu$rad for each monochromatic
component.

The angular spread of x-rays incident on the crystal is $\Delta
x/f_{\ind{1}}=5~\mu$rad, independent of the angular spread of x-rays
incident on the collimating optic. This number is much less than the
crystal angular acceptance, which makes the optic very efficient. The
expected monochromatic image size is $\Delta x^{\prime} \simeq
2.5~\mu$m, which can be resolved by the state-of-the-art position sensitive
x-ray detectors with single photon sensitivity \cite{SBD12}.

A very good energy resolution of $\Delta E \simeq 0.1$~meV
simultaneously requires a very high momentum transfer resolution
$\Delta Q \simeq 0.01$-nm$^{-1}$ to resolve photon-like excitations in
disordered systems (see Fig.~\ref{fig000}). This limits the angular
acceptance on the collimating optic to $\Delta \alpha \lesssim \Delta Q/Q
= 0.21$~mrad, where $Q = 46.28$~nm$^{-1}$ is the momentum of a photon
with energy $E=9.132$~keV. The geometrical aperture of the collimating
optic therefore can be small, $\Delta a_{\ind{1}} = f_{\ind{1}} \Delta
\alpha \simeq 0.2$~mm, assuming $f_{\ind{1}}=1$~m. The geometrical
aperture of the focusing optic should be much larger, because the beam
size increases by a factor of $\Delta \sz^{\prime}/\Delta \sz \simeq
9$ to $\Delta a_{\ind{2}}\simeq 1.8$~mm. However, optics with such
apertures are feasible, in particular if advanced grazing incidence
mirrors are used.  We note also that the cumulative dispersion rate of
the two-crystal dispersing element is
$\fcomm{}=5.63~\mu$rad/meV. Hence, the total angular spread of x-rays
after the dispersion element within the imaging window $\Delta
E_{\ind{\cup}}=47$~meV is $\Delta \theta_{\ind{\cup}}^{\prime}\simeq
266~\mu$rad, which can be totally captured by the state-of-the-art
mirrors.

It should be noted that focusing is required only in one dimension,
like for the spectrographs discussed in Section~\ref{cdw-uhrix}.  This
property can be used to simultaneously image the spectrum of x-rays
along the $x$-axis and the momentum transfer distribution along the
$y$-axis (see Fig.~\ref{fig011}), using a 2D position sensitive
detector.

The spectrograph with the two-crystal (+--)-type dispersing element
introduced in the present section has almost an order-of-magnitude
broader spectral window of imaging compared to that of the
spectrograph with the CDW dispersing element, as discussed in
Section~\ref{cdw-uhrix}. However, its realization requires a focusing
mirror with larger geometric aperture and larger focal distance
$f_{\indrm{2}}$.

\begin{figure}[t!]

\setlength{\unitlength}{\textwidth}
\begin{picture}(1,1.04)(0,0)
\put(0.0,0.00){\includegraphics[width=0.5\textwidth]{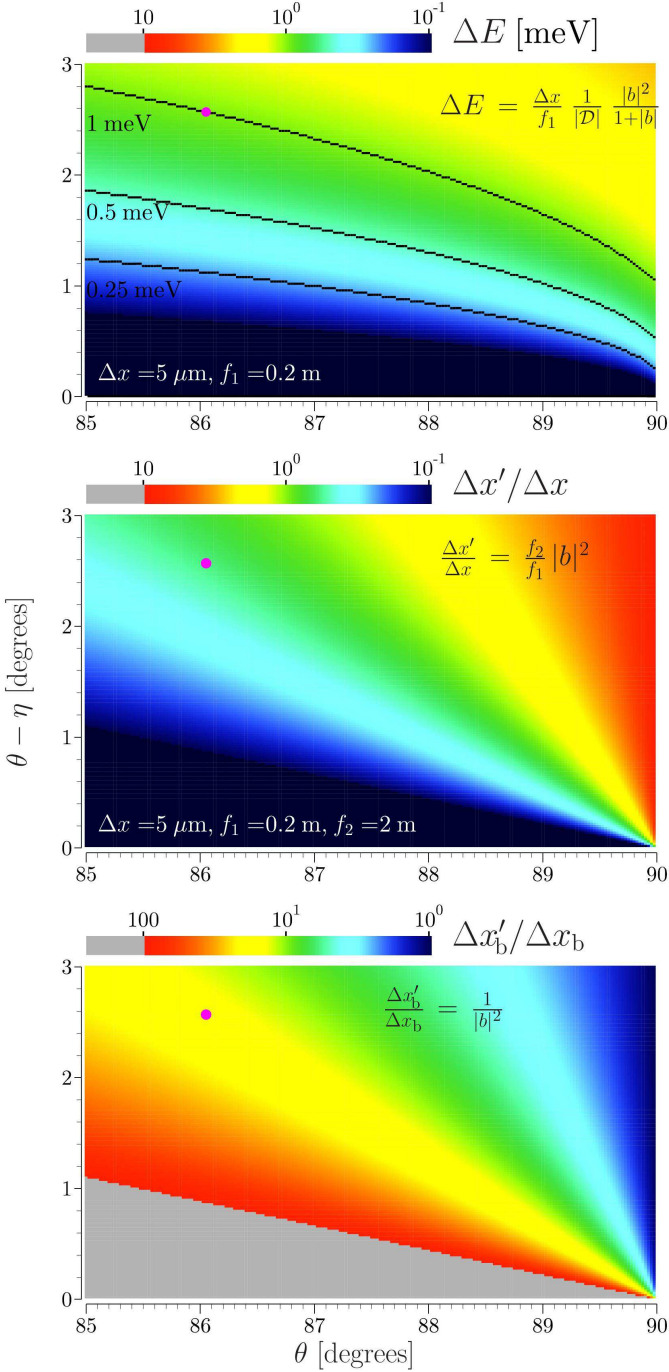}}
\end{picture}
\caption{(Color online) Properties of a spectrograph for Cu K-edge RIXS
  applications. Notations are the same as in Fig.~\ref{fig009} for a
  particular case of $\Delta x=5~\mu$m, $f_{\ind{1}}=0.2$~m, and
  $f_{\ind{2}}=2$~m.  The (337) Bragg reflection of the $8.99$-keV
  x-rays from Ge crystals with crystal parameters presented in
  Table~\ref{tab3}(b) provide a ``diffraction grating'' featuring a
  133-$\mu$rad angular acceptance and a 85-meV bandwidth. In this case,
  the spectrograph resolution should be $\Delta E=1$~meV, $\Delta
  x^{\prime}/\Delta x \simeq 0.3$, and $\Delta \sz^{\prime}/\Delta \sz
  \simeq 16$. }
\label{fig009899}
\end{figure}

\subsubsection{RIXS spectrograph:  1-meV resolution and 85-meV spectral window}
\label{two-cristal-rixs}

Having the Bragg angle $\theta$ as close as possible to $90^{\circ}$
is advantageous, because this allows for better spectral resolution
[see Eq.~\eqref{ssr060}] and simultaneously smaller beam size
enlargement, $\Delta \sz^{\prime}/\Delta \sz$, by the dispersing
element, and not too much reduction of the image to source size ratio
$\Delta x^{\prime}/\Delta x$. This property was used in the example of
the spectrograph intended for IXS applications discussed in the
previous section (see Fig.~\ref{fig009}).  RIXS, unlike IXS, requires
specific photon energies, which are defined by transitions between
specific atomic states \cite{AVD11}. As a consequence, there is
usually limited flexibility in the choice of Bragg's angle
magnitude. Here, we show that in such conditions high-resolution hard
x-ray spectrographs in the \CT\ configuration are also feasible, yet,
with certain limitations.

As an example, we consider a spectrograph for Cu K-edge RIXS
applications, which requires x-rays with photon energies $E\simeq
8.99$~keV. Figure~\ref{fig008}(b) shows a schematic of the
spectrograph's two-crystal (+--)-type dispersing element;
Fig.~\ref{fig009899} displays properties of the spectrograph as a
function of Bragg $\theta$ and asymmetry $\eta$ angles.  Magenta dots
highlight a specific configuration that results in a $\Delta
E=1$~meV spectral resolution, and a $\Delta E_{\ind{\cup}}=85$~meV
spectral window of imaging.  Table~\ref{tab3}(b) presents crystal
parameters in this configuration.

The spectral resolution of the selected RIXS spectrograph is an order
of magnitude inferior to that of the IXS spectrograph discussed in the
previous section. The $\Delta E=1$-meV value is first of all a
compromise between as small as possible spectral resolution and a beam
cross-section that is not overly enlarged by the dispersing
element. In our case the enlargement is already significant: $\Delta
\sz^{\prime}/\Delta \sz \simeq 16$ and will require focusing optic
with large geometric aperture. An overly large deviation of the
$86^{\circ}$ Bragg angle from $90^{\circ}$ (imposed by Ge crystal
properties and fixed photon energy) does not allow for smaller beam
size \footnote{A one-dimensional focusing x-ray mirror can be made with a large
  geometrical aperture by stacking mirror segments
  \cite{UA84}.}. Second, to ensure the larger angular acceptance of the
spectrograph important for RIXS applications, the focal distance of
the collimating optic, which is critical for better spectral
resolution [see Eqs.~\eqref{ssr030} and \eqref{ssr060}] is chosen,
$f_{\ind{1}}=0.2$~m, much smaller that in the IXS case.

The RIXS spectrograph introduced here features an order-of-magnitude
better spectral resolution compared to the resolution available with
the state-of-the-art RIXS spectrometers \cite{HAV06,YAS07,MERIX}.
Such high resolution could be useful in studying collective
excitations in condensed matter systems in various fields, primarily
in high-$T_{\indrm{C}}$ superconductors.

\section{Conclusions}

We have developed a theory of hard x-ray \CT -type spectrographs using
Bragg reflecting crystals in multi-crystal arrangements as dispersing
elements. Using the ray-transfer matrix technique, spectral resolution
and other performance characteristics of spectrographs are calculated
as a function of the physical parameters of the constituent optical
elements.  The dynamical theory of x-ray diffraction in crystals is
applied to calculate spectral windows of imaging.

Several optical designs of hard x-ray spectrographs with broadband
spectral windows of imaging are proposed and their performance is
analyzed. Specifically, spectrographs with an energy resolution of
$\Delta E = 0.1$~meV are shown to be feasible for IXS spectroscopy
applications. Dispersing elements based on CDW optic may provide
spectral windows of imaging, $\Delta E_{\ind{\cup}} \simeq 2.5-7.5$~meV
and compact optical design. Two-crystal (+--)-type dispersing elements
may provide much larger spectral windows of imaging $\Delta
E_{\ind{\cup}} \simeq 45$~meV. However, this may require focusing
optic with a large geometrical aperture, and a large focal length.  In
another example, a spectrograph with a 1-meV spectral resolution and
$\simeq 85$-meV spectral window of imaging is introduced for Cu K-edge
RIXS applications.

Ray-transfer matrices derived in the paper for optics comprising
focusing, collimating, and multiple Bragg-reflecting crystal elements
can be used for the analysis of other x-ray optical systems, including
synchrotron radiation beamline optics, or x-ray free-electron laser
oscillator cavities \cite{KSR08,KS09,Shv13}.\\

\begin{centering}
{\bf ACKNOWLEDGMENTS}\\
\end{centering}
Work at Argonne National Laboratory was supported by the
U.S. Department of Energy, Office of Science, under Contract
No. DE-AC02-06CH11357.



\end{document}